\documentclass[aps,twocolumn,showpacs]{revtex4}
\usepackage{epsf}
\begin{document}
\title{Thermodynamics of Lattice Heteropolymers}
\author{Michael Bachmann}
\email[E-mail: ]{michael.bachmann@itp.uni-leipzig.de}
\author{Wolfhard Janke}
\email[E-mail: ]{wolfhard.janke@itp.uni-leipzig.de}
\homepage[\\ Homepage: ]{http://www.physik.uni-leipzig.de/CQT}
\affiliation{Institut f\"ur Theoretische Physik, Universit\"at Leipzig,
Augustusplatz 10/11, D-04109 Leipzig, Germany}
\begin{abstract}
We calculate thermodynamic quantities of HP lattice proteins by means
of a multicanonical chain growth algorithm that connects the new variants of the 
Pruned-Enriched Rosenbluth Method (nPERM) and flat histogram sampling of the 
entire energy space. Since our method directly simulates the density of states,
we obtain results for thermodynamic quantities of the system for all
temperatures. In particular, this algorithm enables us to accurately simulate the 
usually difficult accessible low-temperature region. Therefore, it becomes possible
to perform detailed analyses of the low-temperature transition between ground states 
and compact globules.    
\end{abstract}
\maketitle
\section{Introduction}
\label{intro}
The native conformation of a protein is strongly correlated with the
sequence of amino acid residues building up the heteropolymer. The
sequence makes the protein unique and assigns it a specific function
within a biological organism. The reason is that the different types
of amino acids vary in their response to the environment and in
their mutual interaction. It is a challenging task
to reveal on what general principles the folding process of a protein 
is based. Models differ extremely in their level of abstraction, ranging
from simple and purely qualitative lattice models to highly sophisticated
all-atom off-lattice formulations with explicit solvent that partially yield 
results comparable with experimental data. Due to the enormous
computational effort required for simulations of realistic proteins,
usually characteristic properties of a protein with a given sequence 
are studied in detail. Much simpler, but by no means trivial, lattice
models enjoy a growing interest, since they allow a more global view on,
for example, the analysis of the relation between sequence and 
structure.    

In this paper, we shall focus ourselves on thermodynamic properties of lattice proteins
at all temperatures. In particular, this includes the investigation of 
the transitions between the different classes of states: lowest-energy 
states, compact globules, and random coils. Since the ground-state--globule
transition occurs at rather low temperatures, a powerful algorithm
is required that in particular allows a reasonable sampling of the low-lying
energy states. To this end we combined multicanonical strategies~\cite{muca1,muca2,hist}
with chain growth algorithms~\cite{grassberger1,grassberger2,hsu1,hsu2}
to a new method~\cite{bj1} which works temperature-independent and directly simulates
the density of states. This quantity contains all energetic information
necessary for computing the mean energy, free energy, entropy, and specific heat
for all temperatures. In the following,
we present results obtained from the application of this method to different 
lattice proteins with lengths up to 103 monomers, modelled by
the simplest lattice formulation
for heteropolymers, the HP model~\cite{dill1}. In this model, only two types of monomers 
enter, hydrophobic (H) and polar (P) residues. The model is
based on the assumption that the hydrophobic interaction is one of the fundamental
principles in protein folding. An attractive hydrophobic interaction
provides for the formation of a compact hydrophobic core that is screened
from the aqueous environment by a shell of polar residues. Therefore the
energy function reads
\begin{equation}
\label{hpmodel}
E = -\sum\limits_{\langle i,j<i-1\rangle} \sigma_i\sigma_j,
\end{equation}
where $\langle i,j<i-1\rangle$ denotes summation over nearest lattice 
neighbours that are nonadjacent along the self-avoiding chain of monomers. A hydrophobic monomer
at the $i$th position in the chain has $\sigma_i = 1$ and a polar monomer is assigned $\sigma_i = 0$.

The first part of this paper, where we explain how our new method works, will be of a more 
technical nature, while the second part is devoted to applications of this algorithm 
to heteropolymers. In Section~\ref{msvscg} we discuss the main differences of Monte Carlo
methods based on move sets for updating conformations and chain growth algorithms
as well as their peculiarities. This is followed by Section~\ref{thermo} about the thermodynamic quantities
that will be estimated with our method. Then, in Section~\ref{mucacg}, we enter into the description
of the multicanonical chain growth algorithm. This technical part is preluded by recalling the
essential ingredients of PERM and nPERM$^{\rm ss}_{\rm is}$~\cite{grassberger1,grassberger2,hsu1,hsu2} 
as these are fundamental for setting
up our algorithm. Then we proceed with the explanation of multicanonical chain growth and the
determination of the multicanonical weight factors. Section~\ref{valid} is devoted to the
validation of our method, and in Section~\ref{results} we present the results 
obtained with our algorithm.
There we focus on thermodynamic properties of heteropolymers with sequences of more than 
40 monomers. Finally, we summarise the main aspects of this paper in Section~\ref{summary}.  
\section{Move Sets vs.\ Chain Growth}
\label{msvscg}
Polymers fold on a lattice into conformations that are by definition self-avoiding.
This takes into account the finite volume and the uniqueness of the monomers. A lattice
site can hence be occupied by a single monomer only. This has the consequence that
the number of very dense conformations of a polymer is by orders of magnitude lower 
than that of random coil states. In Monte Carlo simulations, particular attention
must therefore be devoted to efficient update procedures which also allow the sampling of
dense conformations. In polymer simulations, so-called move sets were applied with some
success to study the behaviour near the $\Theta$ point, which denotes the phase
transition, where polymers subject to an attractive interaction collapse 
from random coils to compact conformations.
Move sets being widely used usually consist of transformations that change the 
position of a single monomer and a single bond vector (end flips), a single monomer
position but two bonds (corner flips), two positions and three bonds (crankshaft) or 
moves with more changes, and pivot rotations, where the
$i$th monomer serves as pivot point and one of the two partial chains connected with it
is rotated about any axis through the pivot~\cite{moveset}. 
For (a-thermal) self-avoiding walks the latter method is known to be very efficient~\cite{madsok}.
It becomes inefficient, however, 
the more dense the conformation is. At low temperatures, the acceptance rate of
locally changing a dense conformation decreases drastically and the simulation threatens
to get stuck in a specific conformation or to oscillate between two states. Since the search for 
ground states is an essential aspect of studying lattice proteins, the application of
move sets is not very useful, at least for chains of reasonable length.

A more promising alternative is the completely different approach based on chain growth.  
The polymer grows by attaching the $n$th monomer at a randomly chosen next-neighbour 
site of the $(n-1)$th monomer. The growth is stopped, if the total length $N$
of the chain is reached or the randomly selected continuation of the chain is already
occupied. In both cases, the next chain is started to grow
from the first monomer. This simple chain growth is also not yet very efficient, since 
the number of discarded chains grows exponentially with the chain length. 
The performance can be improved with the Rosenbluth chain growth method~\cite{rosenbluth}, where
first the free next neighbours of the 
$(n-1)$th monomer are determined and then the new monomer is placed to one of 
the unoccupied sites. 
Since the probability of each possibility for the next monomer
to be set varies with the number of free neighbours, this implies a bias given by
\begin{equation}
p_n\sim \left(\prod_{l=2}^n m_l\right)^{-1},
\end{equation}
where $m_l$ is the number of free neighbours to place the $l$th monomer. The bias 
is corrected by assigning a Rosenbluth weight factor $W_n^{\rm R}\sim p_n^{-1}$ 
to each chain that has been generated by this procedure. 
Nevertheless, this method suffers from attrition too: If all next neighbours are 
occupied, i.e., the chain was running into a ``dead end'' (attrition point), the complete chain
has to be discarded and the growth process has to be started anew.  

Combining the Rosenbluth chain growth method with population control, however, as is done 
in PERM (Pruned-Enriched Rosenbluth Method)~\cite{grassberger1,grassberger2}, leads to
a further considerable improvement of the efficiency by increasing the number of successfully
generated chains. This method renders particularly useful for studying the $\Theta$ point 
of polymers, since then the Rosenbluth weights of the statistically relevant chains 
approximately cancel against their Boltzmann probability. The (a-thermal) Rosenbluth 
weight factor $W_n^{\rm R}$ is therefore replaced by 
\begin{eqnarray}
\label{PERMweight}
&&W_n^{\rm PERM}=\prod\limits_{l=2}^n m_l e^{-(E_l-E_{l-1})/k_BT}, \\
&& 2\le n\le N \quad (E_1=0, \quad W_1^{\rm PERM}=1), \nonumber
\end{eqnarray} 
where $T$ is the temperature and $E_l$ is the energy of the partial chain 
${\bf X}_l=({\bf x}_1,\ldots ,{\bf x}_l)$ created with Rosenbluth chain growth.
In PERM, population control works as follows. If a chain has reached length $n$,
its weight $W_n^{\rm PERM}$ is calculated and compared with suitably chosen
upper and lower threshold values, $W_n^>$ and $W_n^<$, respectively. For 
$W_n^{\rm PERM} > W_n^>$, {\em identical} copies are created which grow then
independently. The weight is equally divided among them. If $W_n^{\rm PERM} < W_n^<$,
the chain is pruned with some probability, say 1/2, and in case of survival, its 
weight is doubled. For a value of the weight lying between the thresholds, the
chain is simply continued without enriching or pruning the sample.    

In the recently developed new variants nPERM$^{\rm ss}_{\rm is}$~\cite{hsu1},
the number of copies is not constant and depends on the ratio of the weight 
$W_n^{\rm PERM}$ compared to the upper threshold value $W_n^>$
and the copies are necessarily chosen to be different (the method of selecting
the copies is based on simple sampling (ss) in \mbox{nPERMss} and a kind of 
importance sampling (is) in \mbox{nPERMis}). This proves quite
useful in producing highly compact polymers and therefore these new methods
are very powerful in determining lowest-energy states of lattice proteins. 
\section{Density of States and Thermodynamic Quantities}
\label{thermo}
In order to investigate the thermodynamic properties of lattice proteins 
accurate simulations are necessary. Due to the difficulties with the
update of conformations at low temperatures and a primary interest in 
detecting lowest-energy conformations, only a few results of thermodynamic 
quantities are found in the literature. Nevertheless, the understanding of
the conformational transitions and the dependence of their
sharpness on the sequence is only possible with algorithms that yield
good results for low as well as for high temperatures. Consequently, reasonable
results can only be obtained, if the method allows the sampling of the
entire energy space. 

All energetic statistical quantities of the protein can be expressed by means of the
density of energetic states $g(E)$, since the partition function of a lattice protein 
with given sequence can be written as
\begin{equation}
\label{psumA}
Z = \sum\limits_{\{{\bf x}\}} e^{-\beta E(\{{\bf x}\})} = \sum\limits_i g(E_i)e^{-\beta E_i},
\end{equation}
where $\beta=1/T$ is the inverse of the thermal energy in natural units. The sum in the first
representation runs over all possible realisations of self-avoiding walks on the
lattice, while in the second expression the sum is taken over all energetic
states a lattice protein can adopt. Then, the expectation value of any energetic 
observable $O(E)$ is simply
\begin{equation}
\label{fEmeanA}
\left\langle O(E)\right\rangle(T)=\frac{1}{Z} \sum\limits_i O(E_i) g(E_i)e^{-\beta E_i},
\end{equation}
and the mean energy as the negative logarithmic derivative of $Z$ with respect 
to $\beta$ is given by
\begin{equation}
\label{EmeanA}
\left\langle E\right\rangle(T)=\frac{1}{Z} \sum\limits_i E_i g(E_i)e^{-\beta E_i}.
\end{equation}
With these expressions, the specific heat $C_V=d\langle E\rangle/dT$ obeys
the fluctuation formula
\begin{equation}
\label{CVA}
C_V(T)=\frac{1}{T^2}\left(\langle E^2\rangle-\langle E \rangle^2\right).
\end{equation}
Moreover, knowing $g(E)$, the Helmholtz free energy is obtained from
\begin{equation}
\label{freeEA}
F(T) = -T \ln \sum_i g(E_i)\exp\{-\beta E_i\}
\end{equation}
and the entropy can be calculated as
\begin{equation}
\label{entropyA}
S(T) = \frac{1}{T}\left[\langle E \rangle(T)-F(T)\right].
\end{equation}

In addition, non-energetic structural quantities are of interest for discussing
the compactness of conformations, such as the end-to-end distance
\begin{equation}
\label{eeA}
R_{\rm ee}=|{\bf x}_N-{\bf x}_1|
\end{equation}
and the radius of gyration
\begin{equation}
\label{gyrA}
R_{\rm gyr}=\sqrt{\frac{1}{N}\sum\limits_{l=1}^N \left({\bf x}_l-{\bf x}_0\right)^2},
\end{equation}
where ${\bf x}_0=\sum_{l=1}^N {\bf x}_l/N$ is the centre of mass of the 
conformation (with all monomers having equal mass). For the calculation
of mean values of non-energetic quantities $O$, Eq.~(\ref{fEmeanA}) is replaced   
by the general formula
\begin{equation}
\label{fOmeanA}
\left\langle O\right\rangle(T)=\frac{1}{Z} \sum\limits_{\{{\bf x}\}} O(\{{\bf x}\}) 
e^{-\beta E(\{{\bf x}\})}.
\end{equation}

In the following, we shall focus ourselves on the study of these thermodynamic 
quantities for different HP lattice proteins and
develop an algorithm that allows a direct simulation of the density
of states. 
\section{Multicanonical Chain Growth Algorithm}
\label{mucacg}
Before we describe the idea behind the multicanonical chain growth algorithm and
the iterative determination of the multicanonical weights which are related
to the density of states, we first recall the canonical chain growth variants 
nPERM$^{\rm ss}_{\rm is}$, on which our algorithm builds up.
\subsection{Canonical Chain Growth}
In the original chain growth algorithm PERM~\cite{grassberger1}, the sample
of chains with length $n<N$ is enriched by identical copies if the weight factor (\ref{PERMweight}) 
is bigger than a threshold value $W_n^>$. In order to obey correct 
Boltzmann statistics, the weight is divided among the clones. If, however,
$W_n^{\rm PERM} < W_n^<$, the chain is pruned with, e.g.\  probability $1/2$, requiring the
weight of a surviving chain to be taken twice. Then one attaches a new monomer at a randomly chosen
free next-neighbour site of the previous one. This is done for all chains that still exist and 
the procedure is repeated
until the total chain length $N$ is reached or the growth of a chain was terminated
by a dead end or due to pruning. After all chains created within this tree have grown until 
their end is reached and thus the present so-called {\em tour} 
is finished, a new growth process starts from the first
monomer, i.e., a new tour begins. Having created an appropriate number of chains with length $n$,
they will be canonically distributed at the given temperature $T$. In fact, this is also true
for all partial chains with intermediate lengths $n < N$, but there are strong correlations
between chains with different lengths $n$. 

In the recently proposed new PERM variants nPERM$^{\rm ss}_{\rm is}$ (new PERM with simple/importance 
sampling (ss/is)) a considerable improvement is achieved by creating different copies, i.e., the chains are
identical in $(n-1)$ monomers but have different continuations, instead of completely
identical ones, since identical partial chains usually show up a similar evolution. Because of the different 
continuations, the weights of the copies can differ. Therefore it is not possible to decide about the number
of copies on the basis of a joint weight. The suggestion is to calculate first a predicted weight which
is then compared with the upper threshold $W_n^>$ in order to determine the number of clones.
Another improvement of PERM being followed up since first applications to lattice proteins is
that the threshold values $W_n^>$ and $W_n^<$ are no longer constants, but are dynamically adapted
with regard to the present estimate for the partition sum and to the number of successfully created chains
with length $n$. The partition sum is proportional to the sum over the weight factors of all conformations of 
chains with length $n$, created with a Rosenbluth chain growth method like, for instance, 
nPERM$^{\rm ss}_{\rm is}$:
\begin{equation}
\label{psumC}
Z_n=\frac{1}{M_{\rm tours}} \sum\limits_t W_n^{{\rm nPERM}^{\rm ss}_{\rm is}}({\bf X}_{n,t}).
\end{equation}
Here, ${\bf X}_{n,t}$ denotes the $t$th generated conformation of length $n$. The proportionality
constant is the inverse of the number of successful tours, i.e., the number of chain growth
starts $M_{\rm tours}$ that led to the generation of at least one chain of length $n$. 
Note that due to this normalisation it is possible to estimate the degeneracy of the energy states.
This is in striking contrast to importance sampling Monte Carlo methods, where the overall constant
on the r.h.s.\ of Eq.~(\ref{psumC}) cannot be determined and hence only {\em relative} degeneracies
can be estimated.

Since \mbox{nPERMss} and \mbox{nPERMis}, respectively, are possible fundamental ingredients for our
algorithm, it is useful to recall in some detail how these chain growth algorithms work. 
The main difference in comparison with the 
original PERM is that, if the sample of chains of length $n-1$ shall be enriched, the 
continuations to an unoccupied next-neighbour site have to be different, i.e., the
weights of these chains with length $n$ can differ. Therefore it is impossible
to calculate a uniform weight like $W_n^{\rm PERM}$ as given in Eq.~(\ref{PERMweight}) 
{\em before} deciding whether to enrich, to prune, or simply to  
continue the current chain of length $n-1$. As proposed in Ref.~\cite{hsu1}, it
is therefore useful to control the population on the basis of a predicted
weight $W_n^{\rm pred}$ which is introduced as
\begin{equation}
\label{Wpred}
W_n^{\rm pred}=W_{n-1}^{{\rm nPERM}^{\rm ss}_{\rm is}}\sum\limits_{\alpha=1}^{m_n}
\chi_\alpha^{{\rm nPERM}^{\rm ss}_{\rm is}},
\end{equation}   
where $m_n$ denotes the number of free neighbouring sites to continue with the $n$th 
monomer. The ``importances'' $\chi_\alpha^{{\rm nPERM}^{\rm ss}_{\rm is}}$ differ for \mbox{nPERMss} and \mbox{nPERMis}. Due to 
its characterisation as a simple sampling algorithm (\mbox{nPERMss}), where all continuations
are equally probable, and as a method with importance sampling (\mbox{nPERMis}), the importances
may be defined as
\begin{eqnarray}
\label{importances}
\chi_\alpha^{\rm nPERMss} &=& 1,\\
\chi_\alpha^{\rm nPERMis} &=& \left(m_n^{(\alpha)} +\frac{1}{2}\right)
e^{-\beta(E_n^{(\alpha)}-E_{n-1})}.
\end{eqnarray}
The expression for \mbox{nPERMis} involves the energy $E_n^{(\alpha)}$ of the choice $\alpha \in [1,m_n]$ for placing the 
$n$th monomer and the number of free neighbours $m_n^{(\alpha)}$ of this choice which is identical with
$m_{n+1}$, provided the $\alpha$th continuation was indeed selected for placing the $n$th monomer.
Since $\chi_\alpha^{\rm nPERMis}$ contains 
informations beyond the $n$th continuation of the chain, \mbox{nPERMis} controls the further growth better than \mbox{nPERMss}.
The predicted weight for the $n$th monomer is now used to decide how the growth of the chain is continued.
If the predicted weight is bigger than the current threshold, $W_n^{\rm pred}>W_n^>$, and $m_n>1$, the 
sample of chains
is enriched and the number of copies $k$ is determined according to the empirical rule 
$k={\rm min}[m_n,{\rm int}(W_n^{\rm pred}/W_n^>)]$. Thus, $2\le k\le m_n$ different continuations will be followed
up. Using \mbox{nPERMss}, the $k$ continuations are chosen randomly with equal probability among the $m_n$
possibilities, while for \mbox{nPERMis}
the probability of selecting a certain $k$-tuple $A=\{\alpha_1,\ldots,\alpha_k\}$ 
of different continuations 
is given by
\begin{equation}
\label{tupleP}
p_A=\frac{\sum\limits_{\alpha\in A}\chi_\alpha}{\sum\limits_{A}\sum\limits_{\alpha\in A} \chi_{\alpha}}.
\end{equation}
Considering the probabilities $p_A$ as partial intervals of certain length, arranging them successively
in the total interval $[0,1]$ (since $\sum_A p_A = 1$), and drawing a random number $r\in [0,1)$, one  
selects the tuple whose interval contains $r$. This tuple of different sites is then chosen to continue
the chain. The corresponding weights are~\cite{hsu1}:
\begin{equation}
\label{nPERMisweights}
W_{n,\alpha_j}^{{\rm nPERM}^{\rm ss}_{\rm is}}=W_{n-1}^{{\rm nPERM}^{\rm ss}_{\rm is}}
\frac{m_n}{k\left(\begin{array}{c}m_n \\ k \end{array}\right) p_A}
e^{-\beta\left(E_n^{(\alpha_j)}-E_{n-1}\right)},
\end{equation}   
where $j\in\{1,\ldots, k\}$ is the index of the $\alpha_j$th continuation within the tuple $A$.
In the special case of simple sampling this expression reduces to 
$W_{n,\alpha_j}^{\rm nPERMss}=W_{n-1}^{\rm nPERMss}m_n \exp[-\beta(E_n^{(\alpha_j)}-E_{n-1})]/k$.
If the predicted weight is less than the lower threshold, $W_n^{\rm pred}<W_n^<$, however,
the growth of this chain is stopped with probability $1/2$. In this case, one traces the chain 
back to the last branching point, where the growth can be continued again, or, if there are 
no branching points, a new tour is started. If the chain survives, the continuation of the
chain follows the same procedure as described above, but now with $k=1$, where Eq.~(\ref{tupleP})
simplifies to
$p_A=\chi_\alpha/\sum_{\alpha} \chi_{\alpha}$ since $A=\{\alpha\}$. In this case
the weight of the chain is taken twice.
For $W_n^< \le W_n^{\rm pred} \le W_n^>$, the chain is continued without enriching or pruning
(once more with $k=1$).

The first tour, where the $n$th monomer is attached for the first time, 
is started with bounds set
to $W_n^>=\infty$ and $W_n^<=0$, thus avoiding enrichment and pruning. For the following tours, we 
use 
\begin{equation}
\label{thresh}
W_n^>=C\frac{Z_n}{Z_1}\frac{c_n^2}{c_1^2},
\end{equation}
where $Z_1\equiv c_1$ is 
the partition function of chains with length unity and $c_i$ the number of created chains with $i$ monomers. 
The constant $C\le 1$ is some positive number and controls the number of successfully generated
chains per tour. For the lower bound we use $W_n^<=0.2 W_n^>$. All these choices are in correspondence to 
Ref.~\cite{hsu1}.   
\subsection{Multicanonical Sampling of Rosenbluth-Weighted Chains}
The idea behind our multicanonical chain growth method is to flatten the canonical
energy distribution provided by nPERM$^{\rm ss}_{\rm is}$. For a given temperature,
the latter algorithms yield accurate canonical distributions over some orders of magnitude.
In order to construct the entire density of states, standard reweighting procedures
may be applied, requiring simulations for different temperatures~\cite{hist}. The low-temperature
distributions are, however, very sensitive against fluctuations of weights which inevitably
occur because the number of energetic states is low, but the weights are high. Thus,
it is difficult to obtain a correct distribution of energetic states, since this requires
a reasonable number of hits of low-energy states. Therefore we assign the chains
an additional weight, the multicanonical weight factor $W_n^{\rm flat}$, chosen such 
that all possible energetic states of a chain of length $n$ possess almost equal
probability of realisation. The first advantage is that states having a low
Boltzmann probability compared to others are hit more frequently. Secondly, the
multicanonical weights introduced in that manner are proportional to the inverse
canonical distribution at temperature $T$, $W_n^{\rm flat}(E)\sim 1/P_n^{{\rm can},T}(E)$, 
respective the inverse density of states 
\begin{equation}
\label{flatWeight}
W_n^{\rm flat}(E)\sim g^{-1}_n(E)
\end{equation}
for $T\to\infty$.
Thus, only one simulation is required and a multi-histogram reweighting is not
necessary. An important conceptual aspect is the fact
that the multicanonical weight factors are unknown in the beginning
and have to be determined iteratively. 

Before we discuss the technical aspects regarding
our method, we first explain it more formally. The energy-dependent multicanonical
weights are trivially introduced into the partition sum~(\ref{psumC}) as suitable 
``decomposition of unity'' in the following
way:
\begin{eqnarray}
\label{psumB}
&&\hspace{-12mm}Z_n=\frac{1}{M_{\rm tours}} \sum\limits_t W_n^{{\rm nPERM}^{\rm ss}_{\rm is}}({\bf X}_{n,t})\nonumber\\
&&\times W_n^{\rm flat}(E({\bf X}_{n,t}))\left[W_n^{\rm flat}(E({\bf X}_{n,t}))\right]^{-1}.
\end{eqnarray} 
Since we are going to simulate at infinite temperature, we express with (\ref{flatWeight}) the partition sum
which then coincides with the total number of all possible conformations as
\begin{equation}
\label{psumBB}
Z_n=\frac{1}{M_{\rm tours}} \sum\limits_t g_n(E({\bf X}_{n,t}))W_n({\bf X}_{n,t})
\end{equation}
with the combined weight 
\begin{equation}
\label{combWeight}
W_n({\bf X}_{n,t})=W_n^{{\rm nPERM}^{\rm ss}_{\rm is}}({\bf X}_{n,t})
W_n^{\rm flat}(E({\bf X}_{n,t})).
\end{equation}
Taking this as the probability for generating chains of length $n$, $p_n\sim W_n$, leads to
the desired flat distribution $H_n(E)$, from which the density of states
is obtained by 
\begin{equation}
\label{gExact}
g_n(E)\sim \frac{H_n(E)}{W_n^{\rm flat}(E)}. 
\end{equation}
The canonical
distribution at {\em any} temperature $T$ is calculated by simply reweighting the
density of states to this temperature, $P_n^{{\rm can},T}(E)\sim g_n(E)\exp\left(-E/T\right)$.
\subsection{Iterative Determination of the Density of States}
In the following, we describe our procedure for the iterative determination of the multicanonical 
weights, from which we obtain an estimate for the density of states. Since there are 
no informations about an appropriate choice for the multicanonical weights in the
beginning, we set them in the zeroeth iteration for all chains $2\le n\le N$ and energies $E$ 
equal to unity, $W_n^{\rm flat,(0)}(E)=1$, and the histograms to be flattened 
are initialised with $H_n^{(0)}(E)=0$. These assumptions render the zeroeth iteration
a pure nPERM$^{\rm ss}_{\rm is}$ run. 

Since we set $\beta=0$ from the beginning, the accumulated histogram of all generated chains of length $n$,
\begin{equation}
\label{histzero}
H_n^{(0)}(E)=\sum\limits_{t}W_{n,t}^{{\rm nPERM}^{\rm ss}_{\rm is}}\,\delta_{E_t\, E},
\end{equation}
is a first estimate of the density of states. In order to obtain a {\em flat} histogram in the next
iteration, we update the multicanonical weights
\begin{equation}
\label{mucaWfirst}
W_n^{{\rm flat},(1)}(E)=\frac{W_n^{{\rm flat},(0)}(E)}{H_n^{(0)}(E)}\quad \forall\ n,E
\end{equation}
and reset the histogram, $H_n^{(1)}(E)=0$. 

The first and all following iterations are multicanonical chain growth runs and proceed along similar lines
as described above, with some modifications. The prediction for the new weight follows again (\ref{Wpred}),
but the importances $\chi_\alpha^{\rm is}$ (\ref{importances}) are in the $i$th iteration introduced as
\begin{equation}
\label{newImportances}
\chi_{\alpha}^{{\rm is},(i)}= \left(m_n^{(\alpha)} +\frac{1}{2}\right)
\frac{W_n^{{\rm flat},(i)}(E_n^{(\alpha)})}{W_{n-1}^{{\rm flat},(i)}(E_{n-1})}.
\end{equation}
In the simple sampling case, we still have $\chi_{\alpha}^{{\rm ss},(i)}=1$. If the sample is enriched
($W_n^{\rm pred}>W_n^>$) the weight~(\ref{nPERMisweights}) of a chain with length $n$ choosing the $\alpha_j$th 
continuation is now replaced by
\begin{equation}
\label{nPERMisNewWeights}
W_{n,\alpha_j}^{\rm ss,is}=W_{n-1}^{\rm ss,is}\,\frac{m_n}{k\left(\begin{array}{c}m_n \\ k \end{array}\right) p_A}\,
\frac{W_n^{{\rm flat},(i)}(E_n^{(\alpha_j)})}{W_{n-1}^{{\rm flat},(i)}(E_{n-1})},
\end{equation} 
where in the simple sampling case (ss) $p_A$ and the binomial factor again cancel each other. 
If $W_n^< \le W_n^{\rm pred} \le W_n^>$, an $n$th possible continuation 
is chosen (selected as described for the enrichment case, but with $k=1$) and the weight
is as in Eq.~(\ref{nPERMisNewWeights}). Assuming that $W_n^{\rm pred}<W_n^<$ 
and that the chain has survived
pruning (as usual with probability $1/2$), we proceed as in the latter case and the chain is assigned twice that weight.
The upper threshold value is now determined in analogy to Eq.~(\ref{thresh}) via 
\begin{equation}
\label{threshFlat}
W_n^>=C\frac{Z_n^{\rm flat}}{Z_1^{\rm flat}}\frac{c_n^2}{c_1^2},
\end{equation}
where 
\begin{equation}
\label{Zflat}
Z_n^{\rm flat}=\sum\limits_{t}W_{n,t}^{\rm ss,is}
\end{equation}
is the estimated partition sum according to the new distribution provided by the weights (\ref{nPERMisNewWeights})
for chains with $n$ monomers and $Z_1^{\rm flat}=Z_1\equiv c_1$. Whenever a new iteration is started,
$Z_n^{\rm flat}$, $c_n$, $W_n^<$ are reset to zero, and $W_n^>$ to infinity (i.e.\ to the upper limit
of the data type used to store this quantity).
If a chain of length $n$ with the energy $E$ was created, the histogram is increased by its weight:
\begin{equation}
\label{histnew}
H_n^{(i)}(E)=\sum\limits_{t}W_{n,t}^{\rm ss,is}\,\delta_{E_t\, E}.
\end{equation} 
From iteration to iteration, this histogram approaches the desired flat distribution $H_n(E)$
and after the final iteration $i=I$, the density of states is estimated by
\begin{equation}
\label{gEstimate}
g_n^{(I)}(E)=\frac{H_n^{(I)}(E)}{W_n^{{\rm flat},(I)}(E)},\quad 2\le n\le N,
\end{equation} 
in analogy to Eq.~(\ref{gExact}). 

In our simulations, we usually performed up to $30$ iterations. The runs $0$ to $I-1$ were
usually terminated after $10^5$--$10^6$ chains of total length $N$ had been produced, while in the
measuring run ($i=I$) usually $10^7$--$10^9$ conformations are sufficient to obtain reasonable statistics.
The parameter $C$ in Eq.~(\ref{threshFlat}) that controls the pruning/enrichment statistics 
and thus how many chains of complete 
length $N$ are generated per tour, was set to $C=0.01$, such that on average $10$ complete chains were 
successfully constructed within each tour. With this choice, the probability for pruning
the current chain or enriching the sample was about $20\%$. In almost all started tours 
$M$ at least one chain achieved its complete length. Thus the ratio between started and successfully 
finished tours $M_{\rm succ}$ is approximately unity, $M_{\rm succ}/M\approx 1$, assuring that
our algorithm performed with quite good efficiency.    

Unlike typical applications of multicanonical or flat histogram algorithms in importance sampling
schemes, where all
energetic states become equally probable such that the dynamics of the simulation corresponds
to a random walk in energy space, the distribution to be flattened in our case is the
histogram that accumulates the weights of the conformations. Hence, if the histogram is flat, a small number 
of high-weighted conformations with low energy $E$ has the same probability as a large number 
of appropriate conformations with energy $E'>E$ carrying usually lower weights.
Therefore the number of actual low-energy hits remains lower than the number of hits of states
with high energy. In order to accumulate enough statistics in the low-energy region, 
the comparative large number
of generated conformations in the measuring run is required. 

We have also implemented
multicanonical chain growth simulations, where we were going to flatten the ``naked'' energy 
distribution, i.e., we tried to equalise the number of hits for all energetic states.
The problem is that this contradicts the philosophy of Rosenbluth chain growth methods,
where the bias connected with the Rosenbluth weight controls the population of samples.
Therefore lowest-energy states were not ``tuned'' by this bias and not hit accordingly.
For applications without special focus to the low-temperature region, it may be, however, an
appropriate alternative to the procedure described above and should be pursued further.  
\section{Validation and Performance}
\label{valid}
Before we discuss the physical results obtained with the multicanonical chain growth algorithm,
we first remark on tests validating the method. We compared the specific heat for very short chains
with data from exact enumeration and found that our method reproduces the exact results with
high accuracy. For a chain with 42 monomers, where exact results are not
available, we performed a multi-histogram reweighting~\cite{ferr1} from canonical distributions at different
temperatures obtained from original \mbox{nPERMis} runs. Here, it turned out that 
our method shows up a considerably higher performance (higher accuracy in spite of lower statistics
at comparable CPU times). 
We also compared with implementations based on sophisticated importance sampling Monte Carlo
schemes, e.g., we have also performed multicanonical sampling~\cite{muca1,muca2} and Wang-Landau 
simulations~\cite{wanglandau} in combination with conformational updates different
from chain growth (e.g.\ move sets as described in Section~\ref{msvscg}). For the present
applications, however, all of these attempts proved to be less efficient.
\subsection{Comparison with Results from Exact Enumeration}
\begin{figure}[t]
\centerline{
\epsfxsize=8.5cm \epsfbox{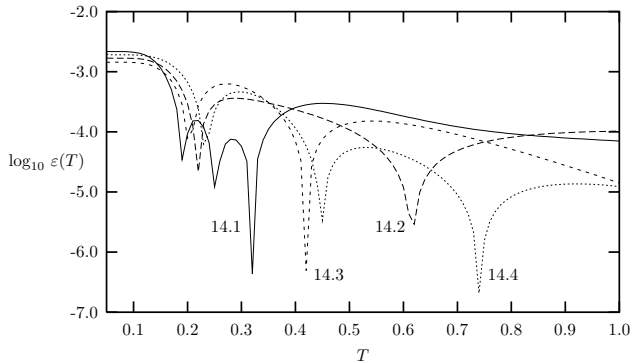}
}
\caption{\label{14comp}
Logarithmic plot of the relative errors of our estimates for the specific heats of 
the 14mers given in Table~\ref{14tab}.
}
\end{figure}

\begin{table}[t]
\caption{
\label{14tab} Sequences, hydrophobicity $n_H$, and global minimum energy $E_{\rm min}$ 
with degeneracy $g_0^{\rm ex}$ (without rotations, reflections, and translations) of the exactly 
enumerated 14mers used for validation of our algorithm. The last column contains the predictions
for the ground-state degeneracy obtained with our method.
}
\begin{tabular}{llcccc}\hline\hline
No. & \multicolumn{1}{c}{sequence} & $n_H$ & $E_{\rm min}$ & $g_0^{\rm ex}$ & $g_0$\\ \hline
14.1 & HPHPH$_2$PHPH$_2$P$_2$H & $8$ & $-8$ & $1$ & $0.98\pm 0.03$\\
14.2 & H$_2$P$_2$HPHPH$_2$PHPH & $8$ & $-8$ & $2$ & $2.00\pm 0.07$ \\
14.3 & H$_2$PHPHP$_2$HPHPH$_2$ & $8$ & $-8$ & $2$ & $2.00\pm 0.06$\\
14.4 & H$_2$PHP$_2$HPHPH$_2$PH & $8$ & $-8$ & $4$ & $3.99\pm 0.13$ \\ \hline \hline
\end{tabular}
\end{table}
As a first validation of our method, we apply it to a set of 14mers with some interesting
properties (see Table~\ref{14tab}) regarding the relation between their ground-state degeneracy and the
strength of the low-temperature conformational transition between lowest-energy states and
compact globules~\cite{bj1,bj2}. In finite-size systems, \mbox{(pseudo-)}transitions are usually identified
through structural peculiarities (maxima for strong transitions or ``shoulders'' for weak transitions) in
the temperature-dependent behaviour of fluctuations of thermodynamic quantities. Usually, it
is hard to obtain a quite accurate estimation of the fluctuations in the low-temperature region.
Thus it is a good test of our method to calculate fluctuating quantities for the 14mers listed in
Table~\ref{14tab} and to compare with results that are still available by exactly enumerating 
all possible $943\,974\,510$ conformations (except translations)~\cite{bj2}. Therefore we determined with
our method the densities of states for these 14mers and calculated the fluctuations
of the energy around the mean value in order to obtain the specific heat according to Eq.~(\ref{CVA}).
We generated $10^9$ chains and the results for the specific heats turned out to be highly accurate.
This is demonstrated in Fig.~\ref{14comp}, where we have plotted for the exemplified 14mers the relative errors 
$\varepsilon(T)=|C_V^{\rm ex}(T)-C_V(T)|/C_V^{\rm ex}(T)$ of our estimates $C_V(T)$ compared with
the specific heats $C_V^{\rm ex}(T)$ obtained by the exact enumeration procedure. We see that, except for
very low temperatures, the relative error is uniformly smaller than $10^{-3}$     
\subsection{Multiple Histogram Reweighting}
\begin{figure}[b]
\centerline{
\epsfxsize=8.5cm \epsfbox{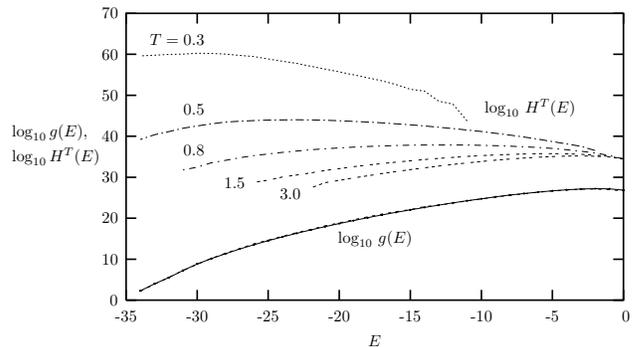}
}
\caption{\label{muHist42}
Histograms $H^T(E)$ obtained by single \mbox{nPERMis} runs for $5$ different temperatures 
$T=0.3$, $0.5$, $0.8$, $1.5$, and $3.0$ (dashed lines).  The resulting density of states 
$g(E)$ obtained by multiple histogram reweighting (long dashed line) lies within the error bars of
the density of states calculated by means of our method (solid line). 
}
\end{figure}
The calculation of the density of states by means of {\em canonical} stochastic algorithms 
cannot be achieved by simply reweighting {\em one} canonical  
histogram, obtained for a given temperature, to as many as necessary distributions to cover
the whole temperature region, since the overlap between the sampled distribution and most of
the reweighted histograms is too small~\cite{hist}. As this simple reweighting only works in a 
certain region around the temperature the simulation was performed, a multiple application of
the reweighting procedure at different sampling temperatures is necessary~\cite{ferr1}.
For the sequence of a 42mer to be studied in detail in Section~\ref{pectate}, 
we performed the multiple reweighting of 5 overlapping
histograms obtained by separate \mbox{nPERMis} runs at temperatures 
$0.3$, $0.5$, $0.8$, $1.5$, and $3.0$ in order to estimate the density of states. 
The histograms as well as the resulting density of states are shown in Fig.~\ref{muHist42},
where we have also plotted the density of states being obtained by means of our
multicanonical sampling algorithm. Each of the histograms contains statistics 
of $8\times 10^7$ chains. This number was adequately chosen such that the density
of states from histogram reweighting matches within the error bars of the density of states 
obtained with our algorithm that also inherently supplies us with the {\em absolute} density
of states. Note that these absolute values cannot be obtained by means of the 
multiple-histogram reweighting procedure, where the normalisation is initially
arbitrary. 
Our density of states was obtained by accumulating statistics of $5\times 10^7$ 
chains. This means that $8$ times more chains were necessary to approximately achieve  
the accuracy with the multiple histogram reweighting method. The iterative period for the 
determination of the multicanonical weights is no drawback, as it takes in our implementation 
only $10\%$ compared to the production run. Therefore we conclude that our dynamical method is 
more efficient and also more elegant than a static reweighting scheme, where also a reliable
estimation of statistical errors is extremely cumbersome. 
\subsection{Flat-Histogram Algorithms with Update Mechanisms Different from Chain Growth}
The calculation of the density of states for heteropolymers with less than 40 monomers
does not represent a big challenge. It is still possible to combine generalised ensemble
methods like multicanonical sampling~\cite{muca1,muca2} or the Wang-Landau method~\cite{wanglandau} 
with move sets 
including pivot rotations to yield reasonable statistics in the low-energy sector.
In order to avoid the sophisticated implementation of the move sets, an alternative
method can be used, for instance, where the conformational information 
is encoded in a string of letters denoting the directions F(orward), B(ackward), R(ight), L(eft), U(p), 
and D(own) the walker may follow in an embedded coordinate system. This structural
sequence is then updated by simply changing a letter. All these methods require
a time-consuming self-avoidance check following each update. 
We have tested these combination of sampling and update methods for 14mers, where
we could compare with exact results from enumeration, and applied it to a 30mer
with 20 energy states with rather high degeneracies. All these states were frequently 
hit such that the results
were reasonable for all temperatures. Remarkably, it turned out, however, that the 
sampling of low-energy states becomes more problematic the lower the degeneracy 
of these states is. Either the algorithm got stuck after hitting such a state,
or it took a long time to find it for the first time. These were indications for
a ``hidden'' conformational barrier that could not be circumvented with these 
procedures. 

Applying these methods to sequences with more than 40 monomers did not
yield reliable results. Low-energy states were too rarely or never hit in long-term
simulations. Performing a biased simulation by explicitly starting from a state
with lowest energy, i.e., initialising with a very dense conformation it took much 
too long until a new self-avoiding conformation was found and accepted.     
Comparing this with applications of the multicanonical chain growth method to these
examples led us to the conclusion that, in the application to lattice proteins,
chain growth methods are much more capable of avoiding such barriers.
\section{Results}
\label{results}
In the following we focus on results which we obtained with the multicanonical chain growth
algorithm for heteropolymers with HP sequences of more than 40 monomers. 
\subsection{Lattice Model for Parallel $\mbox{\boldmath$\beta$}$ Helix with 42 Monomers}
\label{pectate}
We consider a 42mer with the sequence 
PH$_2$PHP\-H$_2$\-P\-H\-P\-H\-P$_2$\-H$_3$\-P\-H\-P\-H$_2$\-P\-H\-P\-H$_3$%
\-P$_2$\-H\-P\-H\-P\-H$_2$\-P\-HPH$_2$P that forms a parallel
helix in the ground state. It was designed to serve as a lattice model of 
the parallel $\beta$ helix of {\em pectate lyase C}~\cite{yoder}. But there are additional
properties that make it an interesting and challenging
system to which we want to apply our method first. The ground-state energy
is known to be $E_{\rm min}=-34$. Moreover, the specific heat has a very 
pronounced low-temperature peak that indicates a \mbox{(pseudo-)}transition between the
lowest-energy states possessing compact hydrophobic cores and the regime of 
the globule conformations. This transition is in addition 
to the usual one between globules and random coils.     
\begin{figure}[t]
\centerline{
\epsfxsize=8.5cm \epsfbox{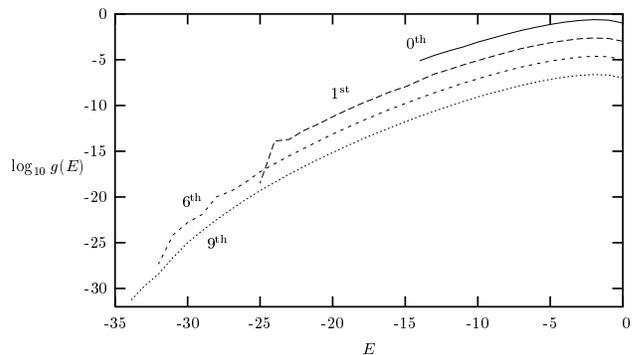}
}
\caption{\label{42iter}
Estimates for the density of states $g_{42}^{(i)}(E)$ for the 42mer
after different levels of iteration.  Since the curves would fall on top of each
other, we have added, for better distinction, a suitable offset to 
the curves of the $1$th, $6$th, and $9$th run. The estimate of the $0$th run is normalised 
to unity.}
\end{figure}
\begin{figure}[t]
\centerline{
\epsfxsize=8.5cm \epsfbox{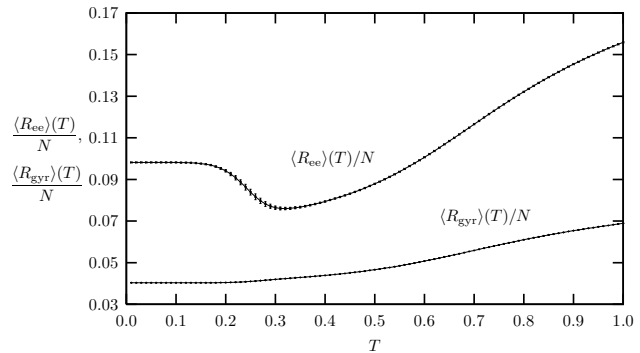}
}
\caption{\label{EG42mer}
Mean end-to-end distance $\langle R_{\rm ee}\rangle$ and mean radius of gyration 
$\langle R_{\rm gyr}\rangle$ of the 42mer.
}
\end{figure}
\begin{figure}[t]
\centerline{
\epsfxsize=8.5cm \epsfbox{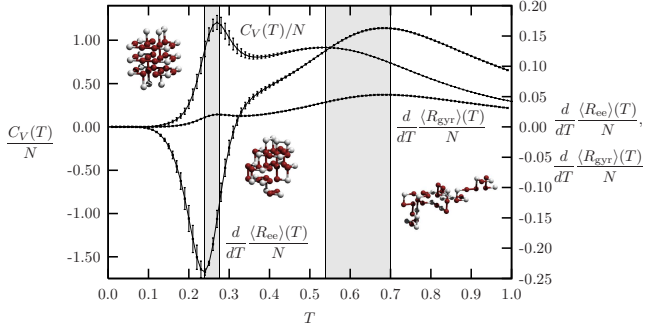}
}
\caption{\label{42mer_fluct}
Specific heat $C_V$ and derivatives w.r.t.\ temperature 
of mean end-to-end distance $\langle R_{\rm ee}\rangle$ and 
radius of gyration $\langle R_{\rm gyr}\rangle$ as functions of temperature for the 42mer. 
The ground-state -- globule transition occurs between $T_0^{(1)}\approx 0.24$ 
and $T_0^{(2)}\approx 0.28$, while the globule -- random coil transition takes
place between $T_1^{(1)}\approx 0.53$ and $T_1^{(2)}\approx 0.70$ (shaded areas). 
}
\end{figure}
\begin{figure}[t]
\centerline{
\epsfxsize=8.5cm \epsfbox{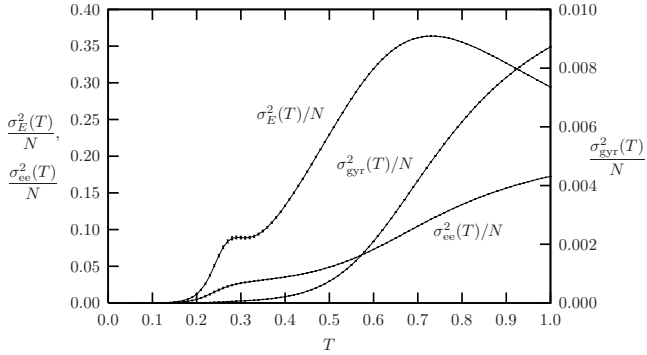}
}
\caption{\label{42mer_fluct2}
Temperature dependence of the fluctuations of energy $\sigma^2_E$, 
end-to-end distance $\sigma^2_{\rm ee}$, and 
radius of gyration $\sigma^2_{\rm gyr}$ for the 42mer. Note the different 
scales for $\sigma^2_{\rm ee}$ and $\sigma^2_{\rm gyr}$.
In the temperature interval plotted the variance of the radius of gyration 
(right scale) is much smaller than the variance of the end-to-end
distance (left scale), $\sigma_{\rm gyr}^2<\sigma_{\rm ee}^2$. 
}
\end{figure}

Figure~\ref{42iter} shows how the estimate for the density of states of this 42mer 
evolves with increasing 
number of iterations. The $0$th iteration is the initial pure
\mbox{nPERMis} run at $\beta=0$. This does not render, however, a proper image of the abilities
of \mbox{nPERMis} which works much better at finite temperatures. Iterations $1$ to $8$
are used to determine the multicanonical weights over the entire energy space
$E\in[-34,0]$. Then, the $9$th iteration is the measuring run which gives a very
accurate estimate for the density of states covering about 25 orders of magnitude. 
Our estimate for the ground-state degeneracy is $g_0=3.9\pm 0.4$, which is in
perfect agreement with the known value $g_0^{\rm ex} = 4$ (except translational,
rotational, and reflection symmetries)~\cite{dill4}. 
\begin{figure}[t]
\centerline{
\epsfxsize=8.5cm \epsfbox{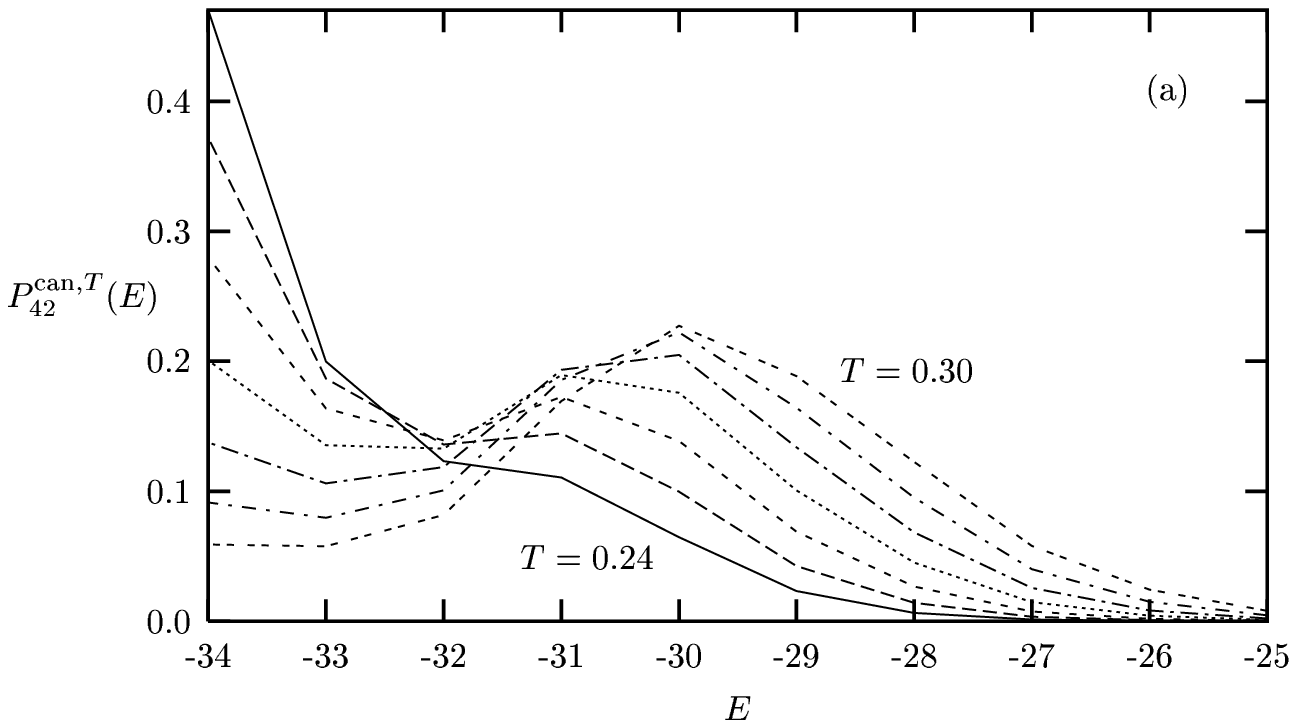}
}
\centerline{
\epsfxsize=8.5cm \epsfbox{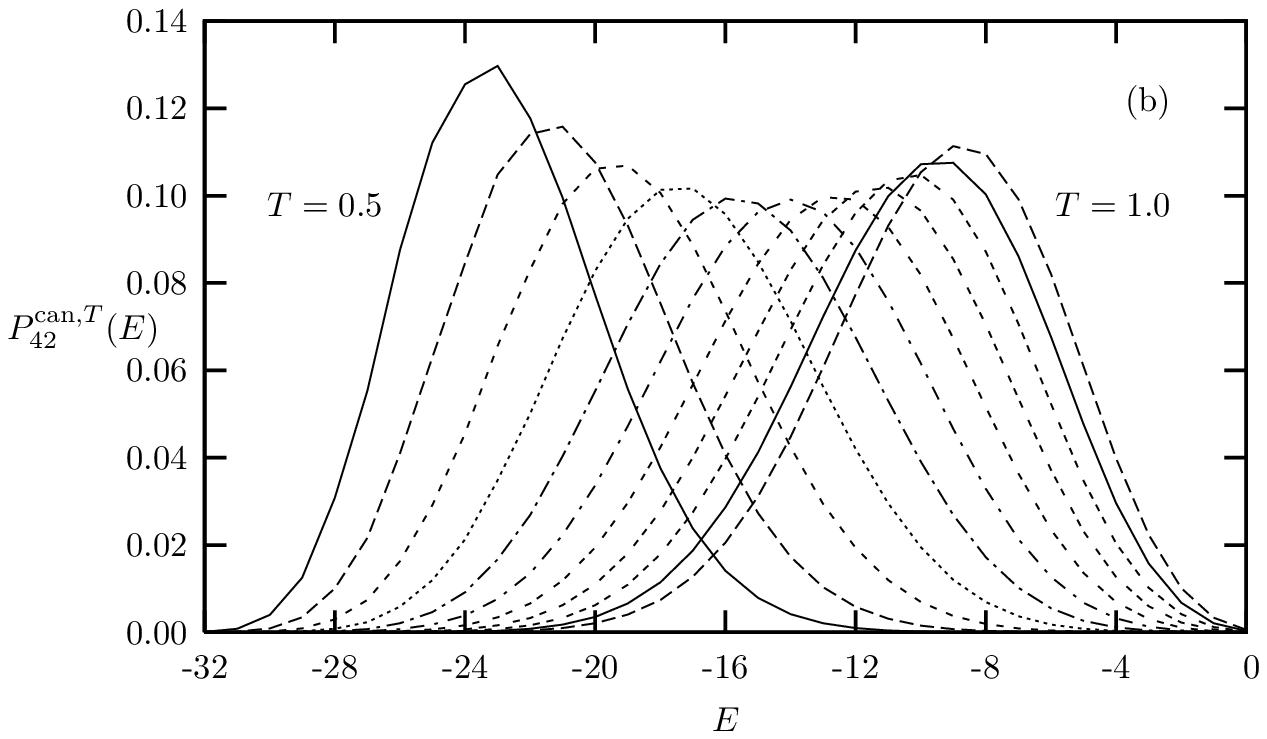}
}
\caption{\label{42mer_pE}
Canonical distributions for the 42mer at temperatures (a) $T=0.24,0.25,\ldots,0.30$ close
to the ground-state -- globule transition region between $T_0^{(1)}\approx 0.24$
and $T_0^{(2)}\approx 0.28$, (b) 
$T=0.50,0.55,\ldots,1.0$. The high-temperature peak of the specific heat in Fig.~\ref{42mer_fluct} 
is near $T_1^{(1)}\approx 0.53$, but 
at $T_1^{(2)}\approx 0.73$ the distribution has the largest width, cf.\ Fig.~\ref{42mer_fluct2}. 
Near this temperature, the
mean radius of gyration and the mean end-to-end distance (see Figs.~\ref{EG42mer} and~\ref{42mer_fluct}) have 
their biggest slope.
}
\end{figure}

The average structural properties at finite temperatures can be best characterised 
by the mean end-to-end distance $\langle R_{\rm ee} \rangle(T)$ and the mean radius 
of gyration $\langle R_{\rm gyr} \rangle(T)$. As these quantities carry shape 
informations, their calculation
is not exclusively based on the density of states $g_N(E({\bf X}_{N,t})$ and hence
Eq.~(\ref{fEmeanA}) cannot be applied.
Therefore expectation values of such quantities $O$ are obtained from the time series of 
the measuring run of the multicanonical chain growth simulation at infinite temperature by
using the general formula
\begin{eqnarray}  
\label{confMean}
&&\hspace{-10mm}\langle O \rangle(T)=\frac{1}{Z_N}\sum\limits_t O({\bf X}_{N,t})\nonumber \\
&&\hspace{10mm}\times W_N({\bf X}_{N,t}) g_N(E({\bf X}_{N,t}))
e^{-\beta E({\bf X}_{N,t})},
\end{eqnarray}
with the estimate for the partition sum 
$Z_N=\sum_t W_N({\bf X}_{N,t})g_N(E({\bf X}_{N,t}))\exp\{-\beta E({\bf X}_{N,t})\}$.
Our results for $\langle R_{\rm ee} \rangle(T)$ and $\langle R_{\rm gyr} \rangle(T)$
of the 42mer are shown in Fig.~\ref{EG42mer}. 
Here and in the following figures, the statistical errors were estimated
by using the jackknife binning method~\cite{jack}.
The pronounced minimum in the end-to-end distance can be interpreted as an indication
of the transition between the lowest-energy states and globules: The
low number of ground states have similar and highly symmetric shapes 
(due to the reflection symmetry of the sequence) but the ends of the chain are 
polar and therefore they are not required to reside close to each other. 
Increasing the temperature allows the protein to fold into conformations different
from the ground states and contacts between the ends become more likely. Therefore, 
the mean end-to-end distance decreases and the protein has entered the globule
``phase''. Further increasing the temperature leads then to a disentangling of the
globules and random coil conformations with larger end-to-end distances dominate.      
From Fig.~\ref{42mer_fluct}, where we have plotted the specific heat and the 
derivatives of the mean end-to-end distance and of the mean radius of gyration 
with respect to the temperature,
\begin{eqnarray}
\label{derivEE}
\frac{d}{dT}\langle R_{\rm ee} \rangle(T) &=& \frac{1}{T^2}\left(\langle ER_{\rm ee}\rangle 
-\langle E\rangle\langle R_{\rm ee}\rangle\right),\\
\label{derivGYR}
\frac{d}{dT}\langle R_{\rm gyr} \rangle(T) &=& \frac{1}{T^2}\left(\langle ER_{\rm gyr}\rangle 
-\langle E\rangle\langle R_{\rm gyr}\rangle\right),
\end{eqnarray}
we estimate the temperature region of the
ground-state~--~globule transition to be within $T_0^{(1)}\approx 0.24$ and 
$T_0^{(2)}\approx 0.28$. The globule~--~random coil transition takes place
between $T_1^{(1)}\approx 0.53$ and $T_1^{(2)}\approx 0.70$. 
\begin{table*}[t]
\caption{\label{48mertab} Properties of the 48mers. For each of the sequences we have 
listed the ground-state energy $E_{\rm min}$ and the ground-state degeneracy $g_0$ estimated
with our algorithm. For comparison, we have also quoted the lower bounds on native degeneracies
$g^<_{\rm CHCC}$ obtained by means of the CHCC (constrained-based hydrophobic core construction) 
method~\cite{dill2} as given in Ref.~\cite{dill3}. In both cases the constant factor $48$ from
rotational and reflection symmetries of conformations spreading into all three spatial directions 
was divided out.}
\begin{tabular}{llrr@{$\, \pm\, $}rr}\hline\hline
No. & \multicolumn{1}{c}{sequence} & $E_{\rm min}$ & \multicolumn{2}{c}{$g_0\ (\times 10^3)$} & 
$g^<_{\rm CHCC}\ (\times 10^3)$  \\ \hline
$48.1$  & HPH$_2$P$_2$H$_4$PH$_3$P$_2$H$_2$P$_2$HPH$_3$PHPH$_2$P$_2$H$_2$P$_3$HP$_8$H$_2$ & $-32$ & 
$5226$ & $812$ & $1500$\hspace{7mm} \\
$48.2$  & H$_4$PH$_2$PH$_5$P$_2$HP$_2$H$_2$P$_2$HP$_6$HP$_2$HP$_3$HP$_2$H$_2$P$_2$H$_3$PH & $-34$ & 
$17$ & $8$ & $14$\hspace{7mm} \\
$48.3$  & PHPH$_2$PH$_6$P$_2$HPHP$_2$HPH$_2$PHPHP$_3$HP$_2$H$_2$P$_2$H$_2$P$_2$HPHP$_2$HP & $-34$ & 
$6.6$ & $2.8$ & $5.0$\hspace{7mm} \\
$48.4$  & PHPH$_2$P$_2$HPH$_3$P$_2$H$_2$PH$_2$P$_3$H$_5$P$_2$HPH$_2$PHPHP$_4$HP$_2$HPHP & $-33$ & 
$60$ & $13$ & $62$\hspace{7mm} \\
$48.5$  & P$_2$HP$_3$HPH$_4$P$_2$H$_4$PH$_2$PH$_3$P$_2$HPHPHP$_2$HP$_6$H$_2$PH$_2$PH & $-32$ & 
$1200$ & $332$ & $54$\hspace{7mm} \\
$48.6$  & H$_3$P$_3$H$_2$PHPH$_2$PH$_2$PH$_2$PHP$_7$HPHP$_2$HP$_3$HP$_2$H$_6$PH & $-32$ & 
$96$ & $19$ & $52$\hspace{7mm} \\
$48.7$  & PHP$_4$HPH$_3$PHPH$_4$PH$_2$PH$_2$P$_3$HPHP$_3$H$_3$P$_2$H$_2$P$_2$H$_2$P$_3$H & $-32$ & 
$58$ & $21$ & $59$\hspace{7mm} \\
$48.8$  & PH$_2$PH$_3$PH$_4$P$_2$H$_3$P$_6$HPH$_2$P$_2$H$_2$PHP$_3$H$_2$PHPHPH$_2$P$_3$ & $-31$ & 
$22201$ & $6594$ & $306$\hspace{7mm} \\
$48.9$  & PHPHP$_4$HPHPHP$_2$HPH$_6$P$_2$H$_3$PHP$_2$HPH$_2$P$_2$HPH$_3$P$_4$H & $-34$ & 
$1.4$ & $0.5$ & $1.0$\hspace{7mm} \\ 
$48.10$ & PH$_2$P$_6$H$_2$P$_3$H$_3$PHP$_2$HPH$_2$P$_2$HP$_2$HP$_2$H$_2$P$_2$H$_7$P$_2$H$_2$ & $-33$ & 
$187$ & $87$ & $188$\hspace{7mm} \\ \hline\hline
\end{tabular}
\end{table*}
\begin{figure}[t]
\centerline{
\epsfxsize=8.5cm \epsfbox{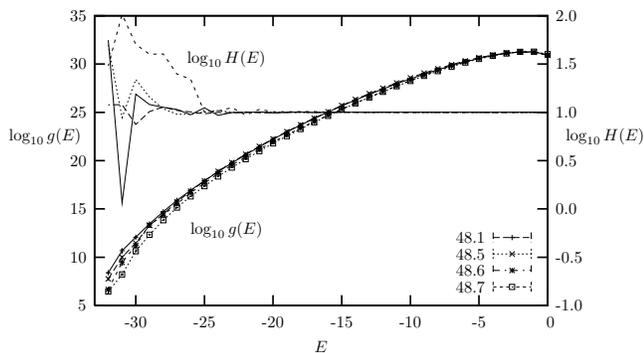}
}
\caption{\label{48mer_gEI}
Logarithmic plots of the densities of states $g(E)$ and ``flat'' 
histograms $H(E)$ for the sequences 48.1, 48.5, 48.6, and 48.7 
from Table~\ref{48mertab} that have the same lowest energy $E_{\rm min} = -32$. 
The normalisation of the histograms of these examples was chosen such that they coincide
at maximum energy, $\log_{10}\,H(E_{\rm max}=0)=1$.
}
\end{figure}

In Fig.~\ref{42mer_fluct2}
we show variances $\sigma^2_O=\langle O^2\rangle-\langle O\rangle^2$ as functions 
of temperature for 
$O=E$, $R_{\rm ee}$, and $R_{\rm gyr}$. We observe weak ``shoulders'' around $T=0.3$ (to
see this for $\sigma^2_{\rm gyr}$ would require, however, an even higher resolution of the plot), close
to the interval, where the low-temperature transition is expected. 
The situation is much more diffuse
in the temperature region, where the globule~--~random coil transition should
take place. The variance of the energy $\sigma_E^2$ has a peak at $T=0.73$, near the
temperatures of the corresponding peaks of the derivatives (\ref{derivEE}) and
(\ref{derivGYR}) plotted in Fig.~\ref{42mer_fluct}. The variances of the end-to-end
distance $\sigma_{\rm ee}^2$ and the radius of gyration $\sigma_{\rm gyr}^2$, 
however, do not exhibit at all a peak near
this temperature. Obviously, there is no unique behaviour of these quantities 
which are usually used to identify conformational transitions in this temperature
region. 
In Fig.~\ref{42mer_pE}
we have plotted the canonical distributions $P_{42}^{{\rm can},T}(E)$ for different
temperatures in the vicinity of the two transitions. From \mbox{Fig.~\ref{42mer_pE}(a)}
we read off that the distributions possess two peaks
at temperatures within that region where the ground-state
-- globule transition takes place. This is interpreted 
as indication of a ``first-order-like'' transition~\cite{iba1}. The behaviour in the
vicinity of the globule -- random 
coil transition is less spectacular as can be seen in \mbox{Fig.~\ref{42mer_pE}(b)}, 
and since the energy distribution 
shows up one peak only, this transition could be denoted as being ``second-order-like''.
The width of the distributions grows with increasing temperature until it has reached its 
maximum value which is located near $T\approx 0.7$, cf.\ Fig.~\ref{42mer_fluct2}. 
For higher temperatures, the distributions become narrower again.   

Since finite-size scaling is impossible because of the non-continuable sequences of
different types of monomers, ``transitions'' between classes of protein
shapes are, of course, to be distinguished from phase transitions in the strict 
thermodynamic sense. 
In conclusion, conformational transitions for 
polymers of finite size, such as proteins, are usually weak and therefore difficult
to identify. Thus, these considerations are, of course, of limited thermodynamic significance.
From a technical point of view, however, it is of some importance
as Markovian Monte Carlo algorithms can show up problems with sampling the 
entire energy space, as the probability in the gap between the two peaks can be suppressed
by many orders of magnitude (what is obviously not the case in our example of the 42mer) 
and tunnelings are extremely rare. Just for such situations, flat histogram algorithms
have primarily been developed~\cite{muca1,muca2}. 
\subsection{Ten Designed 48mers}
\begin{figure}[t]
\centerline{
\epsfxsize=8.5cm \epsfbox{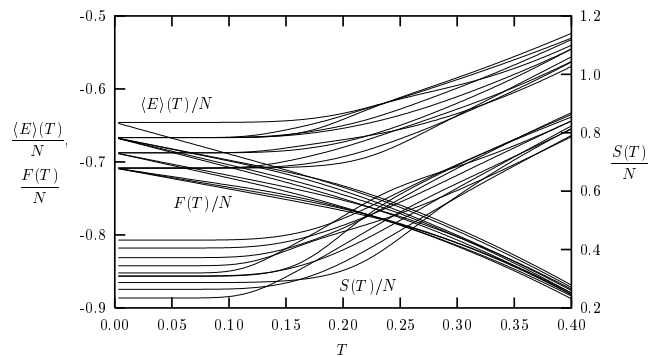}
}
\caption{\label{48mer_EFS}
Mean energy $\langle E\rangle(T)$, Helmholtz free energy $F(T)$, and entropy $S(T)$ for 48mers
with the sequences given in Table~\ref{48mertab}.  
}
\end{figure}
\begin{figure*}
\parbox{8.8cm}{
\epsfxsize=8.5cm \epsfbox{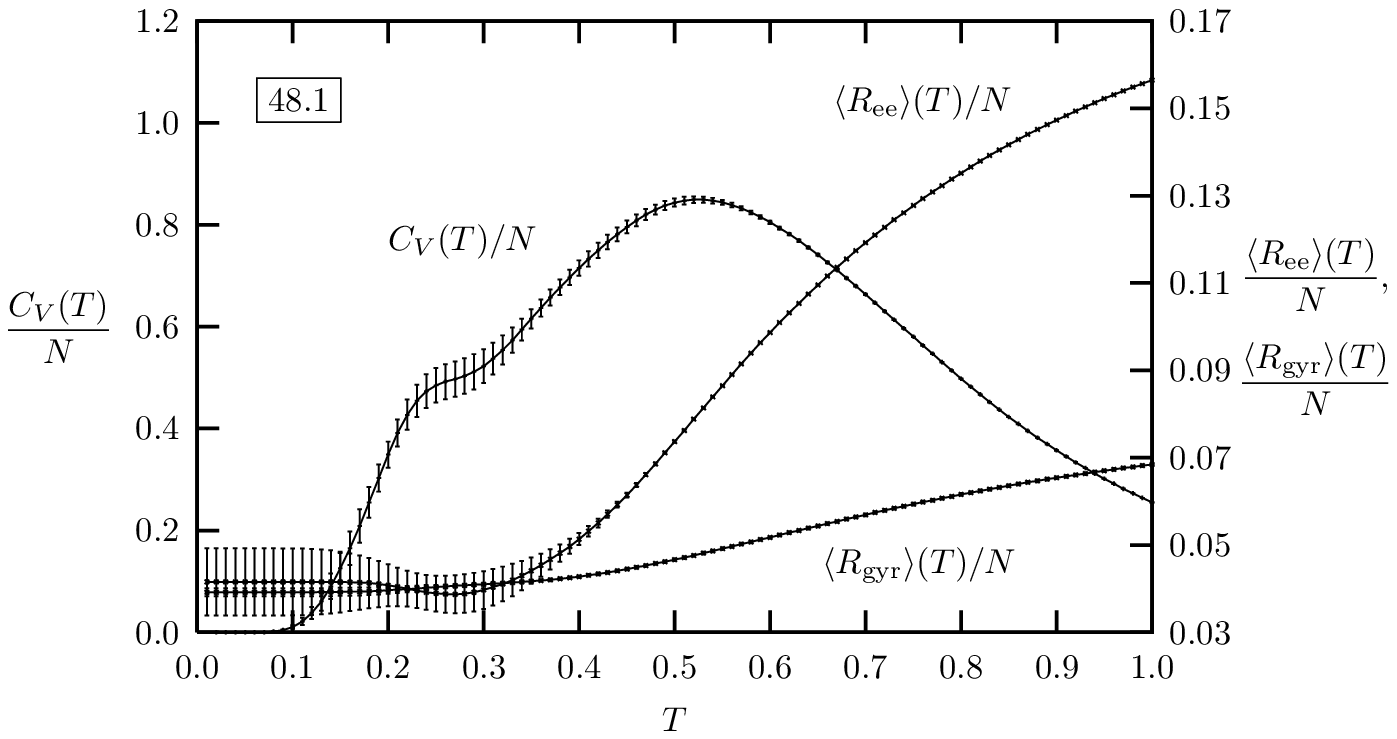}
}\hfill
\parbox{8.8cm}{
\epsfxsize=8.5cm \epsfbox{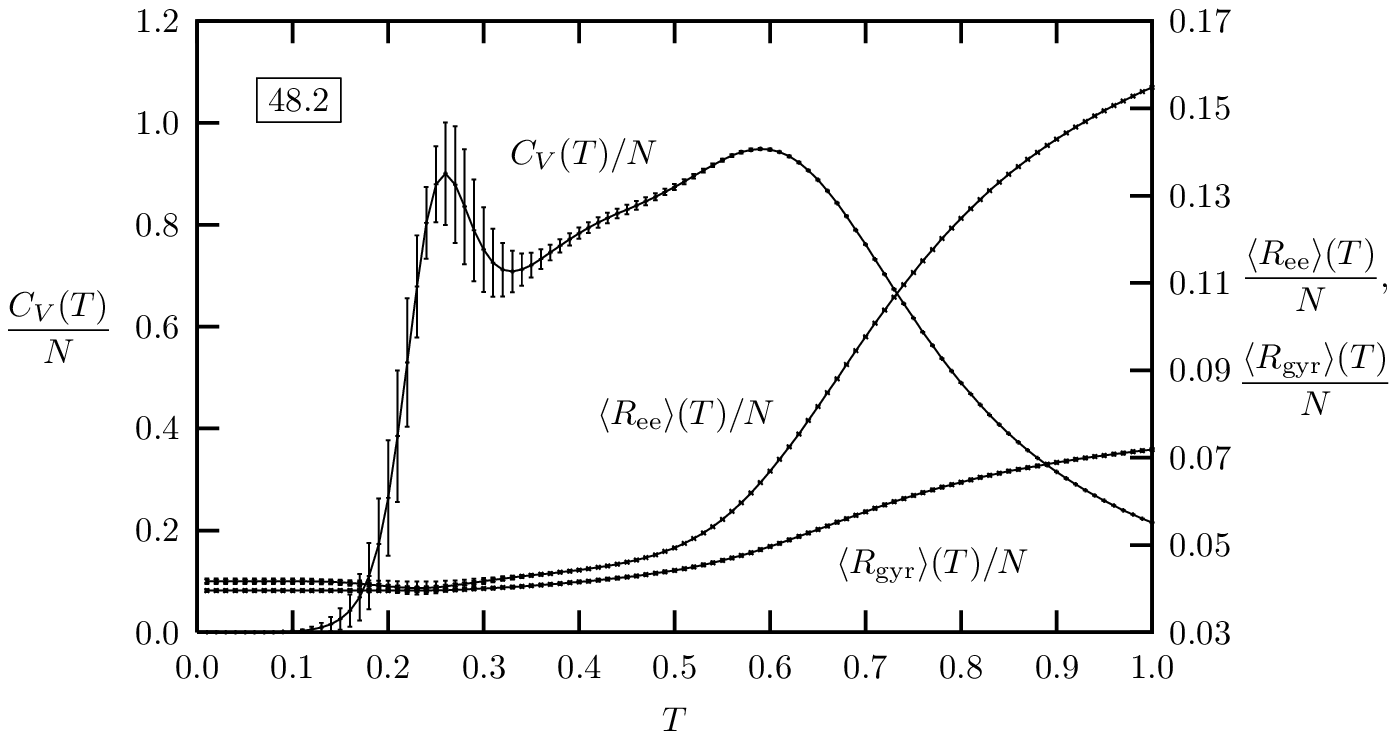}
}\\
\parbox{8.8cm}{
\epsfxsize=8.5cm \epsfbox{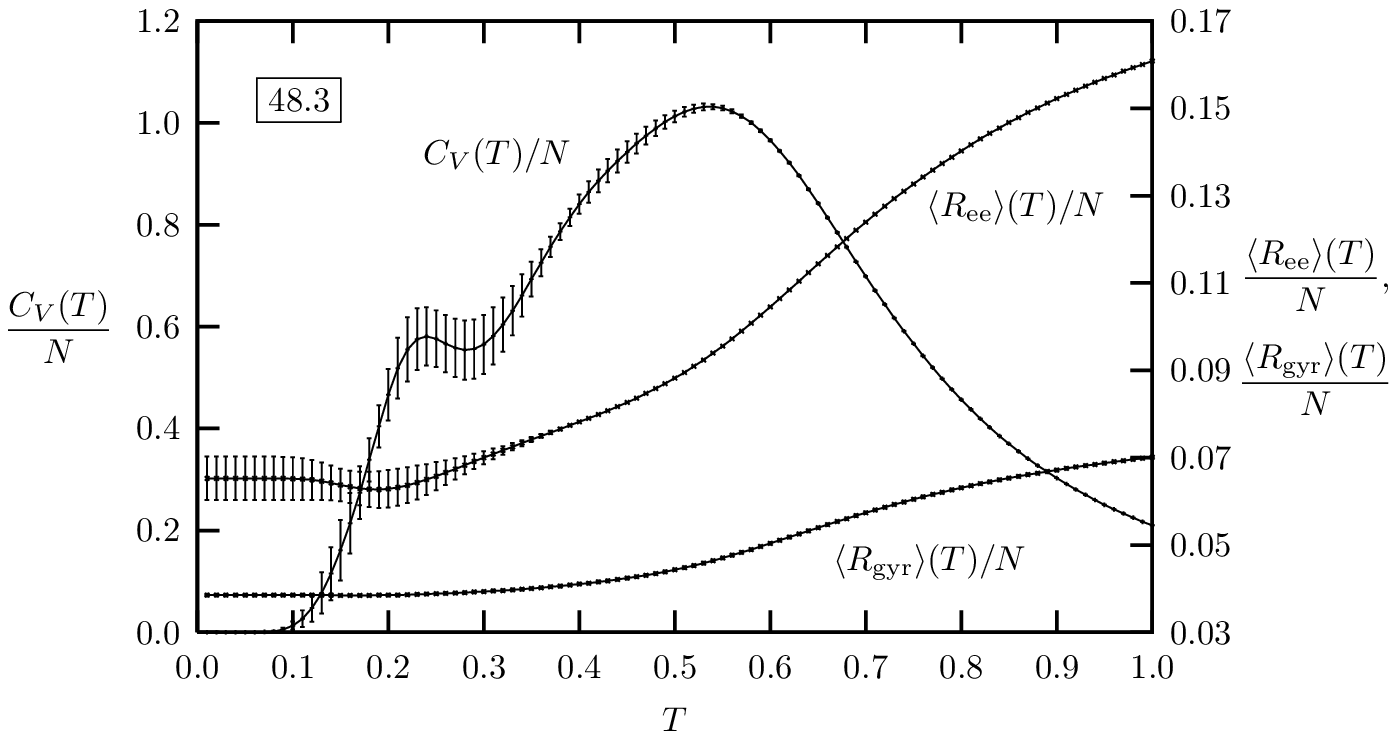}
}\hfill
\parbox{8.8cm}{
\epsfxsize=8.5cm \epsfbox{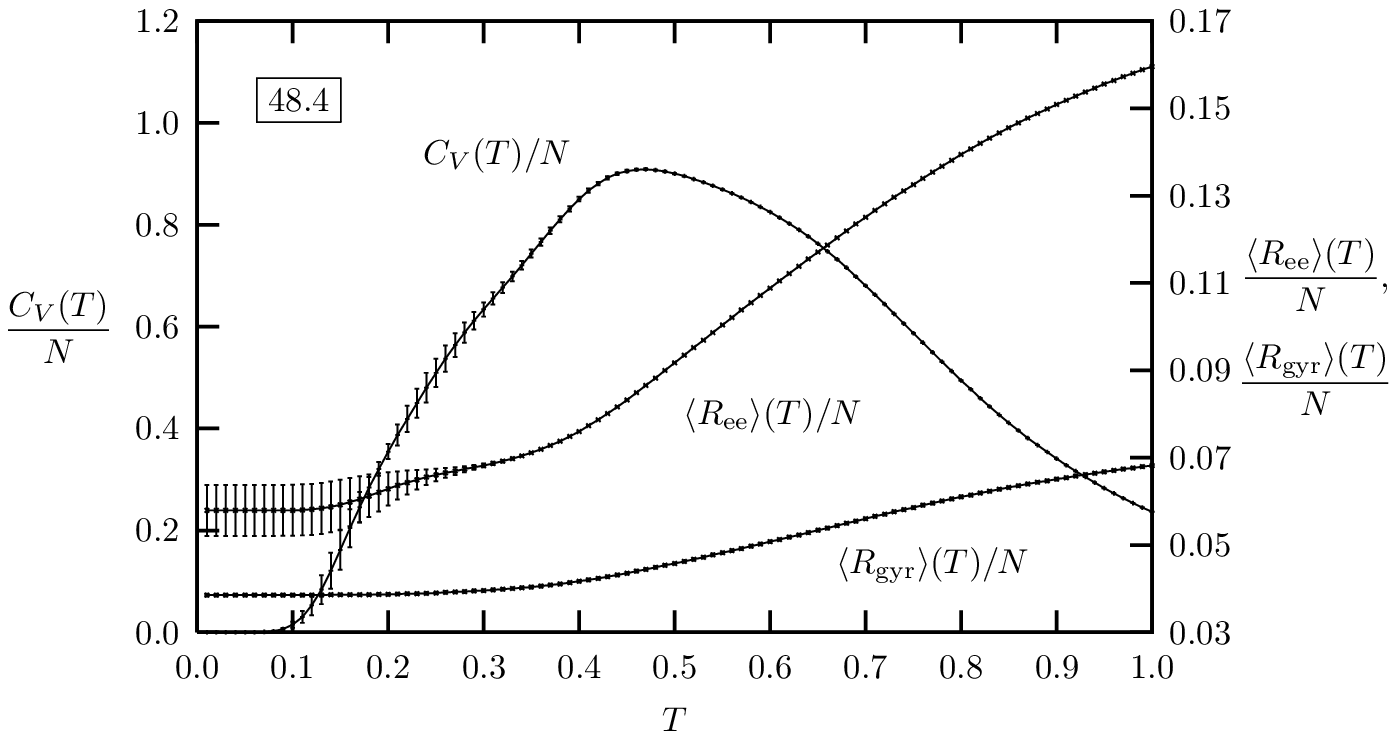}
}\\
\parbox{8.8cm}{
\epsfxsize=8.5cm \epsfbox{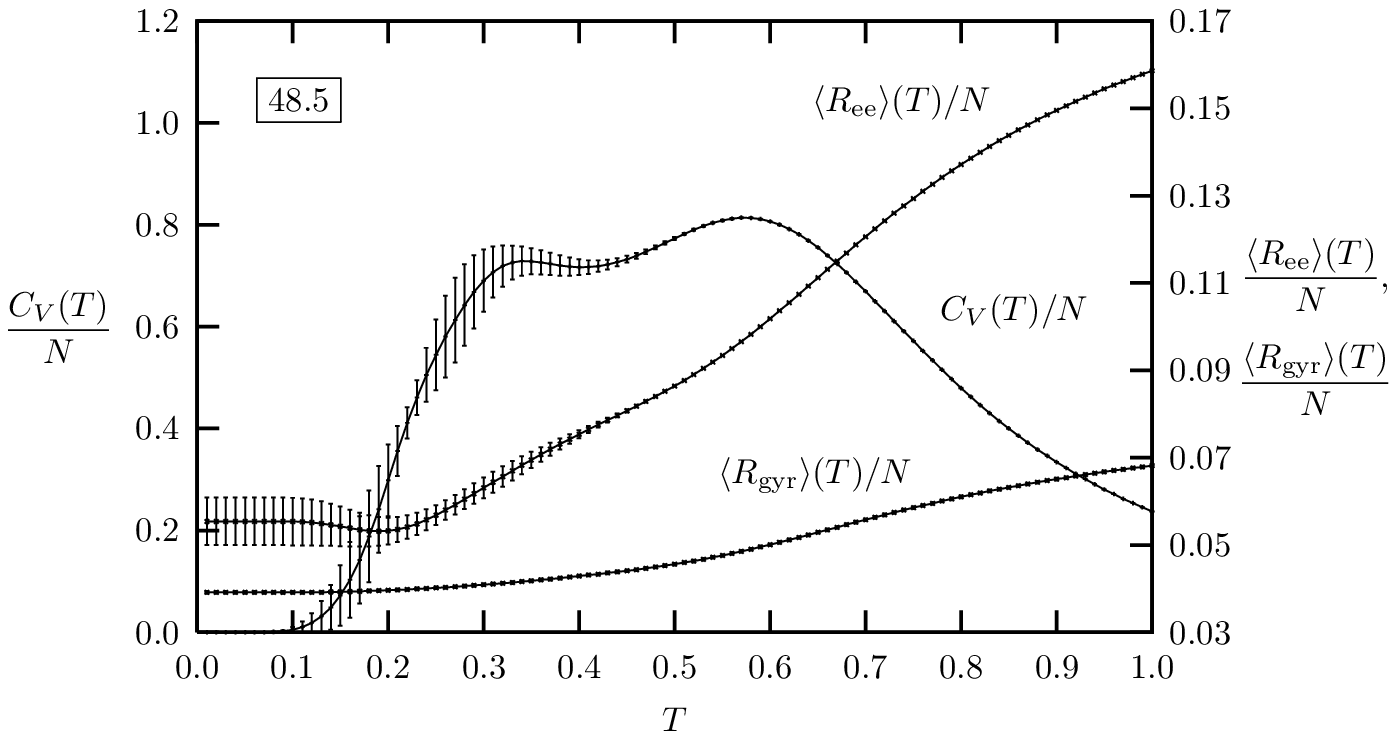}
}\hfill
\parbox{8.8cm}{
\epsfxsize=8.5cm \epsfbox{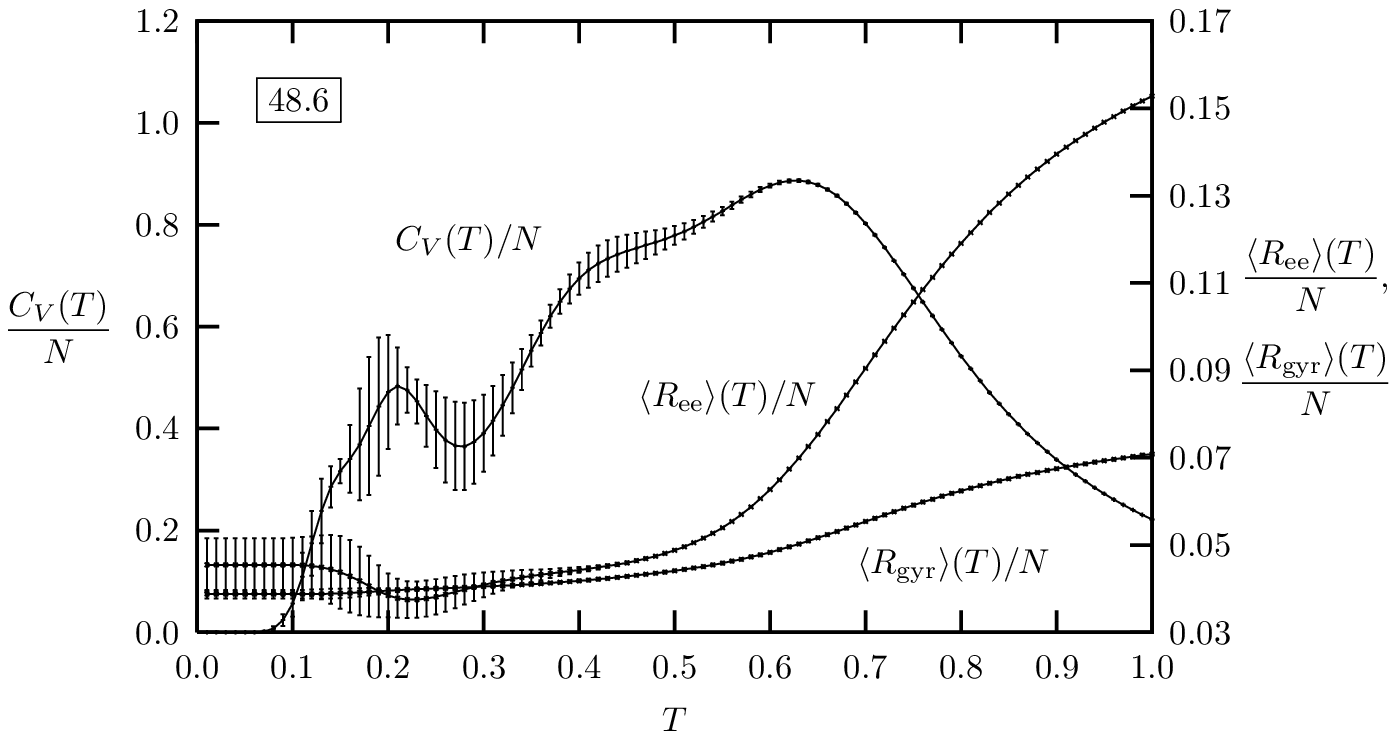}
}\\
\parbox{8.8cm}{
\epsfxsize=8.5cm \epsfbox{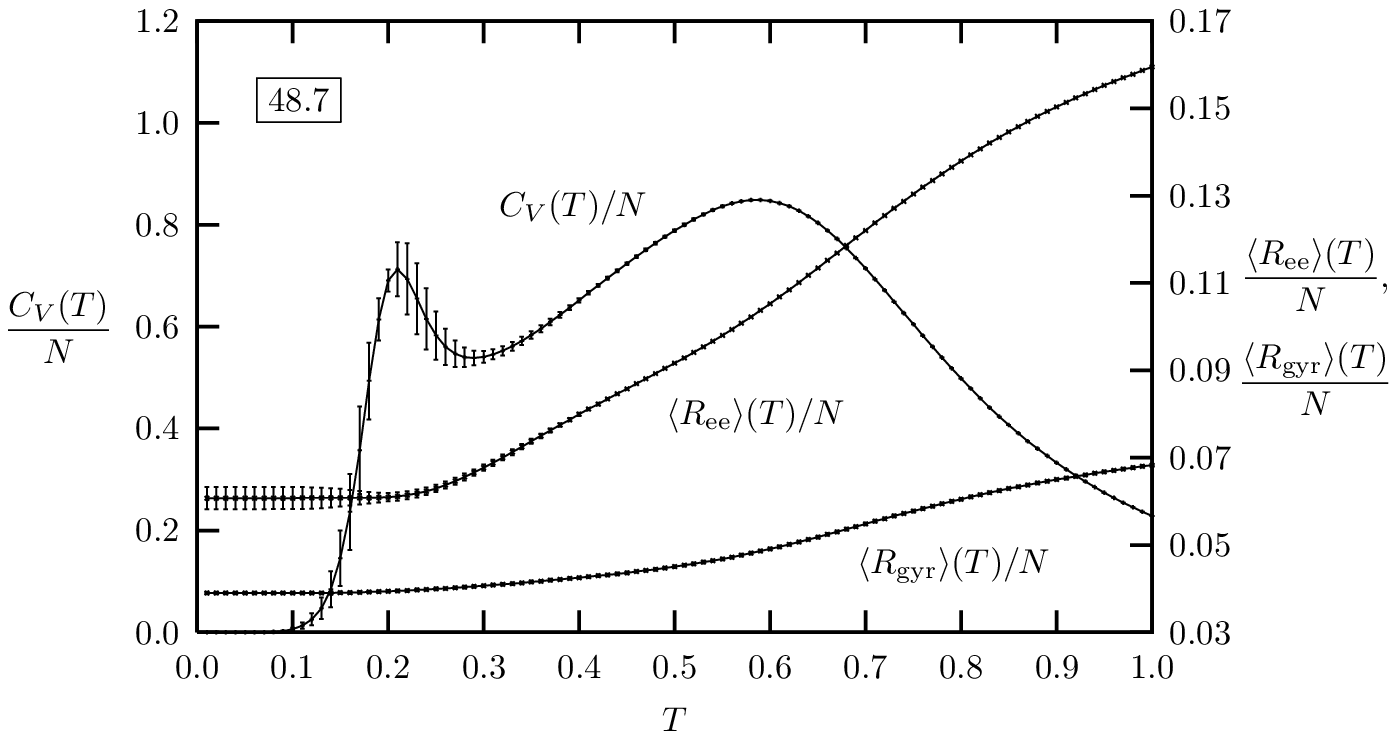}
}\hfill
\parbox{8.8cm}{
\epsfxsize=8.5cm \epsfbox{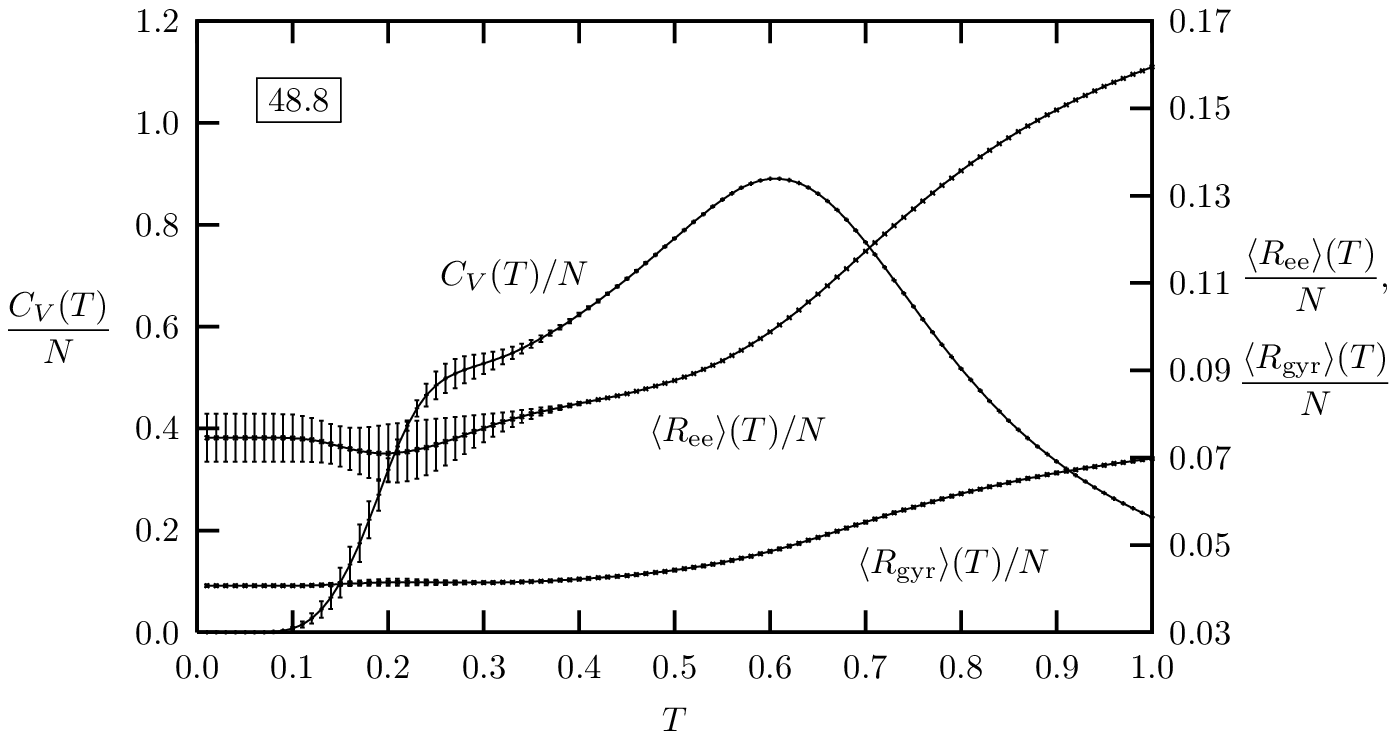}
}\\
\parbox{8.8cm}{
\epsfxsize=8.5cm \epsfbox{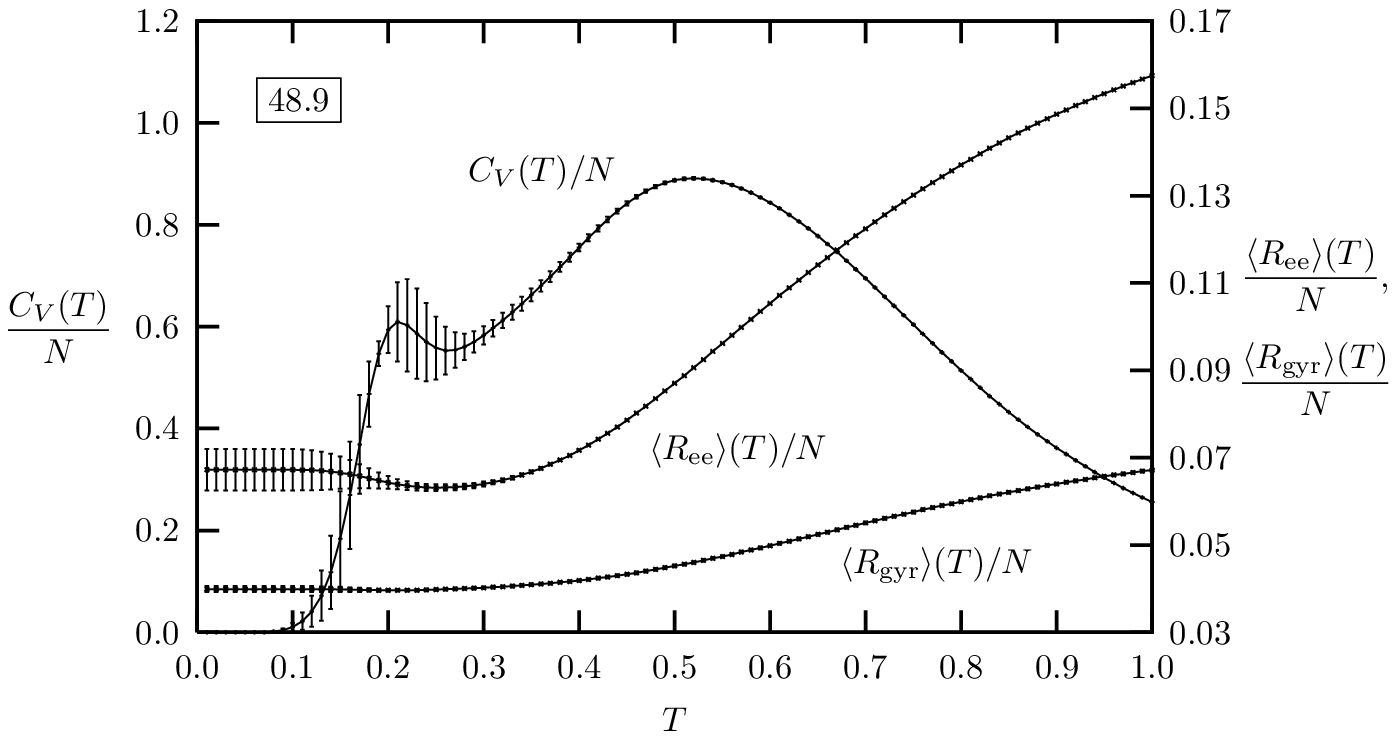}
}\hfill
\parbox{8.8cm}{
\epsfxsize=8.5cm \epsfbox{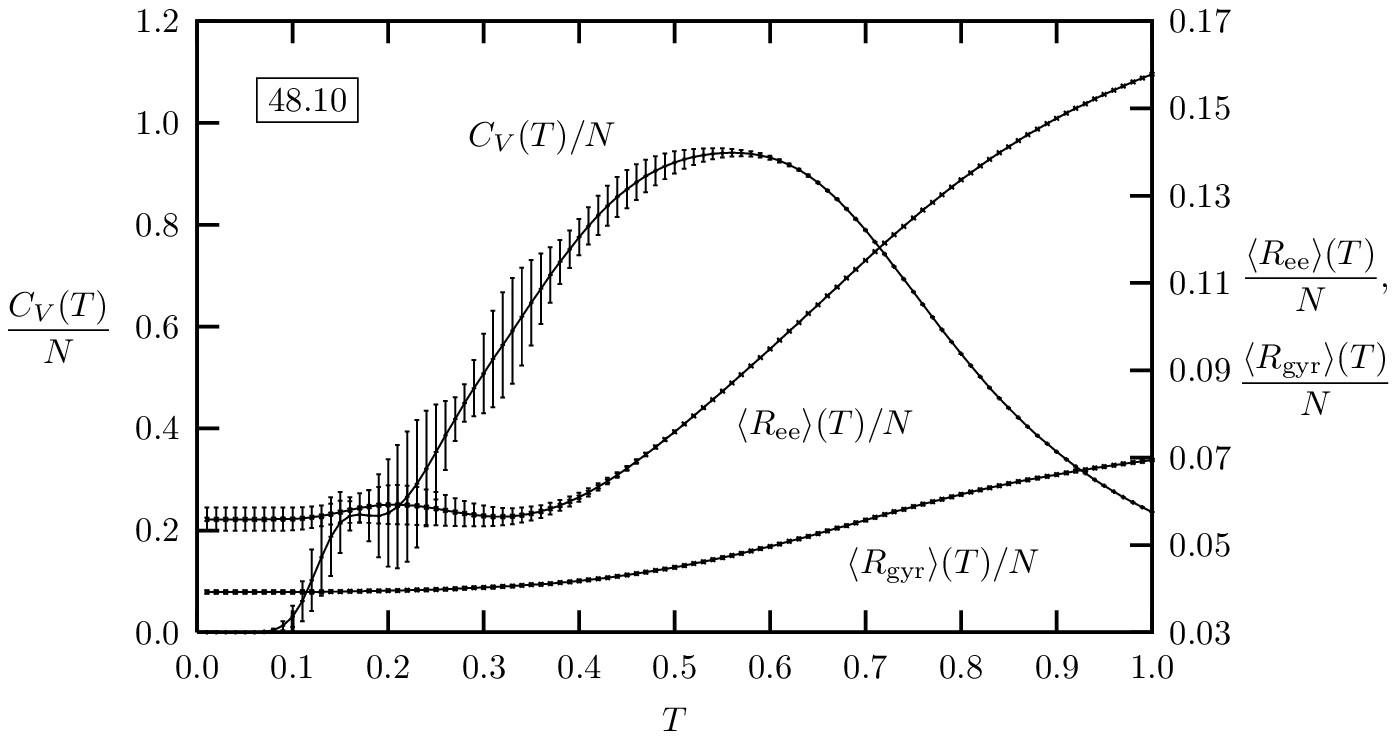}
}
\caption{\label{48mersCR} Heat capacities $C_V(T)$, mean end-to-end distances $\langle R_{\rm ee}\rangle(T)$, and mean radii of gyration $\langle R_{\rm gyr}\rangle(T)$
of the ten designed 48mers from Table~\ref{48mertab}.}
\end{figure*}
We have also analysed the ten designed sequences with 48 monomers given
in Ref.~\cite{dill3}. The ratio between the numbers of hydrophobic and 
polar residues is one half for these HP proteins, i.e., the  hydrophobicity 
is $n_H=24$. In Table~\ref{48mertab} we have listed the sequences and 
ground-state properties. The minimum energies we found coincide with the
values given in Refs.~\cite{dill3,hsu1,hsu2}. 
Figure~\ref{48mer_gEI} shows the densities of states for selected 48mers and the
multicanonical histograms of the production run. 
Note that for Rosenbluth chain growth methods (a-thermal or at $\beta=0$) the 
histogram for chains of length $N$ is obtained by accumulating their individual Rosenbluth 
weights $W_N^{\rm R}$, which explains the poorer performance near the minimum
energy, where a small number of states enters with big weights.  
This differs from the usual procedure in algorithms with 
importance sampling, where the counter of an energy bin being hit by an appropriate 
state is incremented by unity. 

In Fig.~\ref{48mer_EFS} we
have plotted the mean energy, free energy, and entropy as functions
of temperature for these lattice proteins.
These results were obtained by means of the density of states as sampled with our algorithm. 
For $T\to 0$ the curves for both $\langle E\rangle$ and $F$ merge into the
four different values of $E_{\rm min}$ ($=-34$, $-33$, $-32$, $-31$) while the entropy
exhibits the ground-state degeneracies, $S\to {\rm ln}\, g_0$.
Our estimates for the degeneracies $g_0$ of the ground-state energies, and for comparison,
the lower bounds $g^<_{\rm CHCC}$ given in Ref.~\cite{dill3}, are listed 
in Table~\ref{48mertab}. The lower bounds were obtained with
the constrained-based hydrophobic core construction (CHCC) method~\cite{dill2}.   
Our values lie indeed above these lower bounds or include it within the range of statistical
errors. Notice that for the sequences 48.1, 48.5, and 48.8, our estimates for the ground-state 
degeneracy are much higher than the bounds $g^<_{\rm CHCC}$. In these cases the
smallest frame containing the entire hydrophobic core is rather large 
(cube containing $4\times 3\times 3=36$ monomers with surface area 
$A=32$ [bond length]$^2$) such that 
enumeration of this frame is cumbersome. For 48.5 and 48.8, we further found 
ground-state conformations lying in less compact frames (48.5: $A=32,40,42,48,52,54$ [bond length]$^2$, 
48.8: $A=32,40,42$ [bond length]$^2$) and those conformations would require still more
effort to be identified with the CHCC algorithm, which was designed to locate 
global energy minima and therefore starts the search beginning from the most compact 
hydrophobic frames. The ground-state energies of these examples are rather high 
($E_{\rm min}=-31$ for 48.8, and $E_{\rm min}=-32$ for 48.1 and 48.5) and therefore a
higher degeneracy seems to be natural. This is, however, only true, if there does not
exist a conformational barrier that separates the compact H-core low-energy states
from the general compact globules. Comparing the ground-state degeneracies
and the low-temperature behaviour of the specific heats for the sequences 48.1, 48.5, 48.6, and
48.7 (all of them having global energy minima with $E_{\rm min}=-32$) as shown in Figs.~\ref{48mer_gEI} 
and~\ref{48mersCR}, respectively, we observe
that 48.6 and 48.7 with rather low ground-state degeneracy actually possess a pronounced low-temperature 
peak in the specific heat, while the higher-degenerate proteins 48.1 and 48.5 only 
show up a weak indication of a structural transition at low temperatures. The HP proteins
48.2, 48.3, and 48.9, which have the lowest minimum energy $E_{\rm min}=-34$ among the examples
in Table~\ref{48mertab}, have also the lowest ground-state degeneracies. 
These three candidates seem indeed to exhibit a rather strong ground-state -- globule transition, 
as can be read off from the associated specific heats in Fig.~\ref{48mersCR}. 

We have again measured the mean end-to-end distances and mean radii of gyration which are also 
plotted as functions of temperature into Fig.~\ref{48mersCR}. Both quantities 
usually serve to interpret the conformational compactness of polymers. For HP proteins,
the end-to-end distance is strongly influenced, however, by the types of monomers
attached to the ends of the chain. It is easily seen from the figures that the
48mers with sequences starting and ending with a hydrophobic residue (48.1, 48.2, and 48.6)
have a smaller mean end-to-end distance at low temperatures than the other examples from 
Table~\ref{48mertab}. The reason is that the ends can form hydrophobic contacts and therefore
a reduction of the energy can be achieved. Thus, in these cases contacts between ends are 
usually favourable and the mean end-to-end distance is close to the mean radius
of gyration. Interestingly, there exists indeed a crossover region, where 
$\langle R_{\rm ee}\rangle < \langle R_{\rm gyr}\rangle$. Comparing with the behaviour 
of the specific heat, this interval is close to the region, where the 
phase dominated by low-energy states crosses over to the globule-favoured phase. 
The hydrophobic contact between the ends is strong enough to resist the thermal fluctuations
in that temperature interval. The reason is that, once such a hydrophobic contact between the ends is
established, usually other in-chain hydrophobic monomers are attracted and form a
hydrophobic core surrounding the end-to-end contact. Thus, before the contact
between the ends is broken, an increase of the temperature first leads to a melting
of the surrounding contacts. 
The entropic freedom to form new conformations 
is large since the low-energy states are all relatively high degenerate
and do not possess symmetries requiring an appropriate amount of heat to be broken.
For sequences possessing mixed or purely polar ends, the mean end-to-end distance and 
mean radius of gyration differ much stronger, as there is no energetic reason, why 
the ends should be next neighbours. 

In conclusion, we see that for longer chains the strength of the low-temperature transition
not only depends on low ground-state degeneracies as it does for short chains~\cite{bj2}. Rather, 
the influence of the higher-excited states cannot be neglected. A striking example is
sequence 48.4 with rather low ground-state degeneracy, but only weak signals for a 
low-temperature transition.
\begin{figure}[b]
\centerline{
\epsfxsize=5.5cm \epsfbox{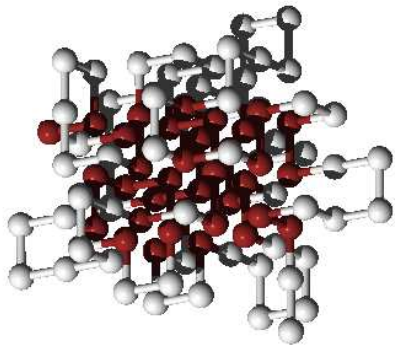}
}
\caption{\label{103mer}
Conformation of the 103mer with the lowest energy found, $E_{\rm min}=-56$.
}
\end{figure}
\subsection{Beyond 100 Monomers ...}
\begin{figure}[t]
\centerline{
\epsfxsize=8.5cm \epsfbox{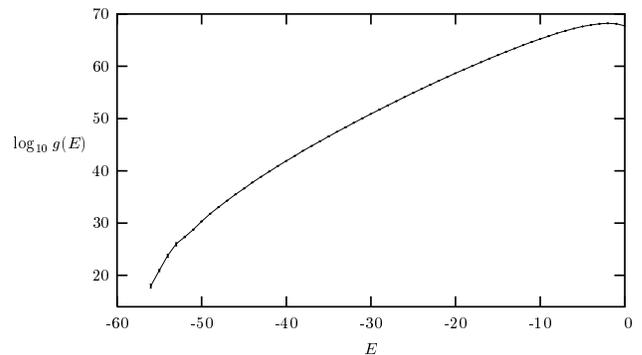}
}
\caption{\label{103mer_gE}
Density of states for the 103mer, ranging over more than 50 orders of magnitude.
}
\end{figure}

\begin{figure}[b]
\centerline{
\epsfxsize=8.5cm \epsfbox{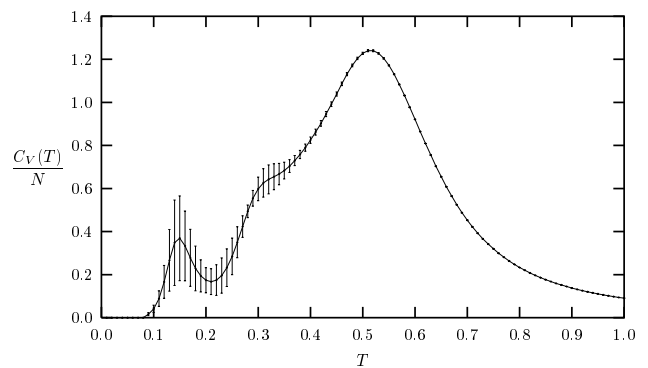}
}
\caption{\label{103mer_CE}
Specific heat of the 103mer.
}
\end{figure}

The final example we applied our algorithm to was a 103mer as
proposed in Ref.~\cite{103lat}. Until recently, the ``ground state''
was believed to have energy $E_{\rm min}=-49$~\cite{103toma}. The up-to-now
best estimate was found with \mbox{nPERMis} to be $E_{\rm min}=-54$~\cite{hsu1} and, with
an additional bias suppressing contacts between H and P monomers, even 
$E_{\rm min}=-55$~\cite{hsu2}. Our algorithm not only decreased the lowest-energy
value to $E_{\rm min}=-56$ (see Fig.~\ref{103mer}), but also enabled us to obtain results for the
thermodynamic quantities as in the previous examples. Figure~\ref{103mer_gE} shows the density
of states which covers more than 50 orders of magnitude from which we determined the specific
heat shown in Fig.~\ref{103mer_CE}. The degeneracy of the lowest-energy state $E_{\rm min}=-56$ was 
determined to be of order $10^{16} $ such that it seems likely that there exist still
one or more even lower-lying energetic states. 
\section{Summary}
\label{summary}
In order to study heteropolymers at very low temperatures with reasonable
accuracy, we developed a multicanonical chain growth algorithm based on
recently improved variants of PERM. Comparing with exact enumeration data
for HP lattice proteins with 14 monomers, we validated that our method 
is suitable to accurately determine thermodynamic quantities for all 
temperatures. Then, we applied this algorithm to lattice proteins with
sequences of more than 40 monomers known from literature. 
Additionally, we determined statistical properties 
for all temperatures for examples with up to 103 monomers. 
Since our algorithm 
allows the estimation of the degeneracy of the energy states, we determined
for all sequences the ground-state degeneracy as it is an indication for
the ``uniqueness'' of the native state. 

In particular, we presented a detailed investigation of a sequence of 42 monomers that has 
interesting characteristic 
properties, e.g., a quite low ground-state degeneracy. Since some results
regarding the ground states and thermodynamic properties were available~\cite{iba1}
it was a good candidate for testing our algorithm and for checking 
the performance of our method. 
For this sequence we analysed in detail the temperature-dependent behaviour 
of radius of gyration, end-to-end distance,
as well as their fluctuations, and compared it with the specific heat in order
to elaborate relations between characteristic properties of these curves 
(peaks, ``shoulders'') and conformational transitions not being transitions
in a strict thermodynamic sense due to the impossibility to formulate
a thermodynamic limit for proteins. Therefore, we identified temperature regions,
where global changes of protein conformations occur. 

Furthermore, we studied energetic and conformational thermodynamic quantities 
for the famous list of ten 48mers given in Ref.~\cite{dill3} in great detail.
These are nice examples as they allow for the comparison of lattice proteins 
with similar properties (same number of monomers, identical hydrophobicity) 
but sequences that differ in the order of hydrophobic and polar monomers.
This also allowed us to study how conformational properties
and the strength of shape transitions depend on the protein sequence. 
As an interesting by-product, we not only confirmed the known global-minimum 
energies for these examples, but we even found a new minimum for the 103mer
being the longest sequence under consideration.  
\section{Acknowledgements}
This work is partially supported by the German-Israel-Foundation (GIF) under
contract No.\ I-653-181.14/1999. 

\begin{thebibliography}{199}
\bibitem{muca1}
B.\ A.\ Berg and T.\ Neuhaus, Phys.\ Lett.\ B {\bf 267}, 249 (1991),
Phys.\ Rev.\ Lett.\ {\bf 68}, 9 (1992).
\bibitem{muca2}
W.\ Janke, Physica A {\bf 254}, 164 (1998); B.\ A.\ Berg, Fields Inst.\ Comm. {\bf 26}, 1 (2000).
\bibitem{hist}
W.\ Janke, {\em Histograms and All That}, in: {\em Computer Simulations of Surfaces and Interfaces}, 
NATO Advanced Study Institute, Albena, Bulgaria, September 2002, edited by D.\ P.\ Landau, A.\ Milchev, 
and B.\ D\"unweg (Kluwer, Dordrecht, 2003).
\bibitem{grassberger1}
P.\ Grassberger, Phys.\ Rev.\ E {\bf 56}, 3682 (1997); H.\ Frauenkron, U.\ Bastolla, E.\ Gerstner, 
P.\ Grassberger, and W.\ Nadler, Phys.\ Rev.\ Lett.\ {\bf 80}, 3149 (1998); U.\ Bastolla, H.\ Frauenkron,
E.\ Gerstner, P.\ Grassberger, and W.\ Nadler, Proteins {\bf 32}, 52 (1998).
\bibitem{grassberger2}
P.\ Grassberger and W.\ Nadler, {\em 'Go with the Winners' Simulations}, in:  
{\em Computational Statistical Physics -- From Billiards to Monte Carlo}, edited by K.\ H.\ Hoffmann and M.\ Schreiber 
(Springer, Berlin, 2002), p.\ 169, and references therein.
\bibitem{hsu1}
H.-P.\ Hsu, V.\ Mehra, W.\ Nadler, and P.\ Grassberger, J.\ Chem.\ Phys.\ {\bf 118}, 444 (2003).
\bibitem{hsu2}
H.-P.\ Hsu, V.\ Mehra, W.\ Nadler, and P.\ Grassberger, Phys.\ Rev.\ E {\bf 68}, 21113 (2003).
\bibitem{bj1}
M.\ Bachmann and W.\ Janke, e-print: cond-mat/0304613, to appear in {\em Phys.\ Rev.\ Lett.} (in print 2003).
\bibitem{dill1}
K.\ A.\ Dill, Biochemistry {\bf 24}, 1501 (1985); K.\ F.\ Lau and K.\ A.\ Dill,
Macromolecules {\bf 22}, 3986 (1989).
\bibitem{moveset}
See, e.g., L.\ W.\ Lee and J.-S.\ Wang, Phys.\ Rev.\ E {\bf 64}, 56112 (2001).
\bibitem{madsok}
N.\ Madras and A.\ D.\ Sokal, J.\ Stat.\ Phys.\ {\bf 50}, 109 (1988).
\bibitem{rosenbluth}
M.\ N.\ Rosenbluth and A.\ W.\ Rosenbluth, J.\ Chem.\ Phys.\ {\bf 23}, 356 (1955).
\bibitem{ferr1}
A.\ M.\ Ferrenberg and R.\ H.\ Swendsen, Phys.\ Rev.\ Lett.\ {\bf 63}, 1195 (1989).
\bibitem{wanglandau}
F.\ Wang and D.\ P.\ Landau, Phys.\ Rev.\ Lett.\ {\bf 86}, 2050 (2001).
\bibitem{bj2}
M.\ Bachmann and W.\ Janke, Acta Physica Polonica B {\bf 34}, 4689 (2003).
\bibitem{yoder}
M.\ D.\ Yoder, N.\ T.\ Keen, and F.\ Jurnak, Science {\bf 260}, 1503 (1993).
\bibitem{dill4}
T.\ C.\ Beutler and K.\ A.\ Dill, Prot.\ Sci.\ {\bf 5}, 2037 (1996).
\bibitem{jack}
R.\ G.\ Miller, Biometrika {\bf 61}, 1 (1974); B.\ Efron, 
{\em The Jackknife, the Bootstrap, and Other Resampling Plans} (Society for Industrial and Applied 
Mathematics [SIAM], Philadelphia, 1982).
\bibitem{iba1}
G.\ Chikenji, M.\ Kikuchi, and Y.\ Iba, Phys.\ Rev.\ Lett.\ {\bf 83}, 1886 (1999), and references therein.
\bibitem{dill3}
K.\ Yue, K.\ M.\ Fiebig, P.\ D.\ Thomas, H.\ S.\ Chan, E.\ I.\ Shaknovich,
and K.\ A.\ Dill, Proc.\ Natl.\ Acad.\ Sci.\ USA {\bf 92}, 325 (1995).
\bibitem{dill2}
K.\ Yue and K.\ A.\ Dill, Phys.\ Rev.\ E {\bf 48}, 2267 (1993); Proc.\ Natl.\ Acad.\ Sci.\ USA
{\bf 92}, 146 (1995).
\bibitem{103lat}
E.\ E.\ Lattman, K.\ M.\ Fiebig, and K.\ A.\ Dill, Biochemistry {\bf 33}, 6158 (1994).
\bibitem{103toma}
L.\ Toma and S.\ Toma, Prot.\ Sci.\ {\bf 5}, 147 (1996).
%
\end{thebibliography}
\end{document}